\documentclass[a4paper,superscriptaddress,11pt]{article}

\title{ Polymer Parametrised Field Theory}
\author{Alok Laddha\\The Institute of Mathematical Sciences,
 Chennai-600 113, India\\alokl@imsc.res.in \and Madhavan Varadarajan\\Raman Research Institute, Bangalore-560 080, India \\madhavan@rri.res.in}

\begin{document}

\maketitle

\begin{abstract}

Free scalar field theory on 2 dimensional flat spacetime, cast in diffeomorphism
invariant guise by treating  the inertial coordinates of the spacetime
as dynamical variables, is quantized using
LQG type `polymer' representations for the matter field and the inertial variables.
The quantum constraints are solved via group averaging techniques and,
analogous to the case of spatial geometry in LQG, the smooth (flat) spacetime
geometry is replaced by a discrete quantum structure.
An overcomplete set of Dirac observables, consisting of (a) (exponentials of)
the standard free scalar field  creation- annihilation modes and (b) canonical
transformations corresponding to conformal isometries,
are represented as operators on the physical Hilbert space.
None of these constructions suffer from any of the `triangulation' dependent
choices which arise in treatments of LQG.
In contrast to the standard Fock quantization,
the non- Fock nature of the representation ensures that the algebra of conformal
isometries as well as that of spacetime diffeomorphisms are represented in an anomaly free manner.
Semiclassical states can be analysed at the
gauge invariant level. It is shown that `physical weaves' necessarily underly such states and that such
states display semiclassicality with respect to, at most, a countable subset of the (uncountably large) set of
observables of type (a).
The model thus offers a fertile testing ground for proposed definitions of quantum dynamics as well
as semiclassical states in LQG.

\end{abstract}

\section{Introduction}
This work is devoted to an application of canonical Loop Quantum Gravity (LQG) techniques to the quantization
of a generally covariant, field theoretic toy model which goes by the name of Parametrised Field Theory (PFT).
PFT is just free field theory on flat spacetime, cast in a diffeomorphism invariant disguise. It offers an elegant
description of free scalar field evolution on {\em arbitrary} (and in general curved) foliations of the
background spacetime by treating the `embedding variables' which describe the foliation as dynamical variables
to be varied in the action in addition to the scalar field. Specifically, let $X^A= (T,X)$ denote inertial coordinates
on 2 dimensional flat spacetime. In PFT, $X^A$ are parametrized by a new set of arbitrary coordinates
 $x^{\alpha}=(t,x)$
such that for fixed $t$, the embedding variables $X^A(t,x)$ define a spacelike Cauchy slice of flat spacetime.
General covariance of PFT ensues from the arbitrary choice of $x^{\alpha}$ and implies that in its canonical
description, evolution from one slice of an arbitrary foliation to another is generated by constraints.
While 2 dimensional PFT has been quantized in a Fock representation for the matter fields in References
\cite{karelqm,charlie}, here we are interested in the construction of an LQG type representation for both the
embedding as well as the matter fields, along the lines of Reference \cite{alok}.
The usefulness of this exercise for canonical LQG can only be gauged  in the context of the current status of
the field, a brief discussion of which we now turn to.

LQG is a non- perturbative approach to quantum gravity which, in its canonical version, attempts to
construct a Dirac quantization of a
Hamiltonian description of gravity in terms of a spatial $SU(2)$ connection.
and its  conjugate electric field. The strength of this approach is that it constitutes, for the most part,
an extremely conservative development and application of canonical quantization techniques to gravity
(see for e.g. the reviews \cite{carloreview,aajlreview,leereview,ttreview}).
This conservative union of the principles of quantum mechanics with those of classical gravity has yielded
many beautiful results such as a satisfactory treatment of spatial diffeomorphisms \cite{RSlooprep,ALM^2T},
discrete spatial geometry \cite{RSarea,aajlarea,jlvolume},
a calculation
of black hole entropy \cite{carloentropy,aaentropy} and a uniqueness theorem for its underlying representation
\cite{LOST,fleischhack}. However, a necessity for radical ideas has arisen in the treatment of quantum dynamics
\cite{ttalqg,spinfoamreview} as well as in that of semiclassical issues \cite{mejose,ttcoherent,aashadow}.

The key obstruction to a completely conservative treatment stems from the fact that in LQG only certain
non- local functionals of the connection, namely the holonomies around spatial loops, can be promoted to
quantum operators rather than the connection itself.
\footnote
{The reason for this is the lack of regularity in the action of the holonomy operators: while, classically,
the connection at a point can be obtained from the holonomy of a loop containing the point in the limit that
the loop is infinitesimally small, the limit of the corresponding operators does not exist in the
LQG representation.
 \label{footnote1}}
As a result, all questions of interest (including that of the quantum dynamics defined by the Hamiltonian constraint
which is a {\em local} function of the connection and triad,) need to be phrased in terms of holonomy operators.
Since holonomy operators associated with
close by loops have actions unrelated by any sort of continuity, this leads to a situation where a {\em choice}
of a subset of the (uncountable) set of all holonomy operators (or equivalently, the spatial loops labelling them)
becomes
necessary. We shall loosely refer to such choices as ``triangulation'' choices since, often, the family of
loops is chosen to lie on some set of triangulations of the spatial manifold. Since there seems to be no
natural choice independent of the intuition of the researcher, this leads to proposals which may be seen as
radical or ad- hoc depending on ones taste.

In order to test these proposals it is necessary  to have a `perfect' toy model in which an LQG type of
quantization can be constructed which is free from any triangulation ambiguities. What is needed is a
generally covariant, {\em field theoretic} (with an infinite number of true degrees of freedom, since many of the
difficulties can be traced to the field theoretic nature of gravity) system in which all steps of an LQG type
quantization procedure can be carried out in a triangulation independent manner.
As we show in this work, just such a model is provided by 2 dimensional parametrised field theory on $S^1 \times R$.
Specifically, we construct, in a triangulation independent manner:
an appropriate kinematic  `holonomy' algebra and its LQG type `polymer' representation on a kinematic Hilbert space
${\cal H}_{kin}$,
a representation on ${\cal H}_{kin}$ of both (the finite transformations generated by) the constraints
and
an over- complete set of gauge invariant observables,
the group averaging map \cite{dongrpavg,ALM^2T} and the physical state space ${\cal H}_{phys}$
which naturally inherits  a representation of the Dirac observables from that on ${\cal H}_{kin}$  .

The above quantization of PFT offers an arena in which proposals for quantum dynamics developed for LQG may be
tested against the manifestly triangulation/regularization free group averaging techniques used in this work.
Further, semiclassical issues can be examined at the physical state level since both ${\cal H}_{phys}$ and
representation of an overcomplete set of Dirac observables thereon, are available. This is in contrast to LQG
wherein most current proposals are defined on ${\cal H}_{kin}$ with the hope that they may still be useful at the
physical state level. Again, since the quantization here admits a representation of  Dirac observables on
${\cal H}_{kin}$ as well as ${\cal H}_{phys}$, it offers a useful testing ground for proposed constructions
of semiclassical states in LQG. Finally, since PFT also admits the usual Fock space quantization of the
scalar field \cite{karelqm,charlie}, this can be compared with the ``polymer'' quantization presented here.
This comparison is useful for similar `graviton from LQG' issues \cite{megraviton} in canonical LQG.

The layout of the paper is as follows. Section 2 contains a brief review of classical PFT on
$S^1\times R$. Details may be found in \cite{karelcm}. In section 3, ${\cal H}_{kin}$ is constructed as the
tensor product of Hilbert spaces for the matter and embedding sectors, each of which supports a polymer representation
of suitably defined LQG- type operators. It is shown that ${\cal H}_{kin}$ also supports a unitary
representation of the finite canonical transformations generated by the constraints. In section 4 an overcomplete set of
gauge invariant (Dirac) observables corresponding to (a) exponentials of the standard mode functions of the free
scalar field on flat spacetime  and  (b) conformal isometries, are promoted to operators on ${\cal H}_{kin}$.
These operators commute with those corresponding to  finite gauge transformations. In section 5, the physical state
space, ${\cal H}_{phys}$, is constructed through group averaging techniques \cite{dongrpavg,ALM^2T}. Ambiguities
in the group averaging map are systematically reduced by requiring commutativity with the Dirac observables and
superselection sectors are described, each of which provide a cyclic, {\em non- separable} representation
of the algebra generated by the gauge invariant operators of section 4. Section 6 is devoted to a preliminary
discussion of semiclassical issues. It is shown that, at most, only a countable subset of the overcomplete
(and uncountable) set of
Dirac observables of type (a) can be approximated by semiclassical states in ${\cal H}_{phys}$.
Further, it is shown that any such state must be characterized by a suitably defined ``physical''
weave. Two issues  (connected with the $S^1$ spatial topology and the treatment of zero modes)
are addressed in section 7.
Section 8 contains a discussion of our results as well as of open issues.

In the interests of brevity, we shall refrain from providing detailed proofs where such proofs are straightforward.
Some Lemmas are proved in the Appendices A and B.  The dimensions of various quantities and our choice of units
 are displayed in Appendix C.

\section{Classical PFT on $S^1\times R$.}
We provide a brief review of classical 2 dimensional PFT. In  sections 2.1 and 2.2  we shall implicitly assume that the
spatial topology is that of a circle. The consequences of this non- trivial spatial topology on the formalism
will be made explicit in section 2.3.

\subsection{The Action for PFT.}

The action for a free scalar field $f$ on
a fixed flat 2 dimensional spacetime in terms of global inertial coordinates $X^A,\;A=0,1$ is
\begin{equation}
S_0[f] = -\frac{1}{2} \int d^{2}X \eta^{AB}\partial_Af \partial_Bf ,
\label{s0}
\end{equation}
where  the Minkowski metric
in inertial coordinates, $\eta^{AB}$,  is diagonal with entries $(-1,1)$. If instead, we use coordinates
$x^{\alpha}\;, \alpha= 0,1$ (so that $X^A$ are `parameterized' by $x^{\alpha}$,  $X^A= X^A(x)$), we have
\begin{equation}
S_0[f] = -\frac{1}{2} \int d^{2}x \sqrt{\eta} \eta^{\alpha \beta}\partial_{\alpha}f \partial_{\beta}f ,
\label{s0x}
\end{equation}
where $\eta_{\alpha \beta}= \eta_{AB} \partial_{\alpha}X^A\partial_{\beta}X^B$ and $\eta$ denotes the
determinant of $\eta_{\alpha \beta}$. The action for PFT is obtained
by considering the right hand side of (\ref{s0x}) as a functional, not only of $\phi$, but also of
$X^A(x)$ i.e. $X^A(x)$ are considered as  2 new scalar fields to be varied in the action ($\eta_{\alpha \beta}$
is a function of $X^A(x)$). Thus
\begin{equation}
S_{PFT} [f, X^A]= -\frac{1}{2} \int d^{2}x \sqrt{\eta(X)} \eta^{\alpha \beta}(X)\partial_{\alpha}f \partial_{\beta}f .
\label{spft}
\end{equation}
Note that $S_{PFT}$ is a diffeomorphism invariant functional of the scalar fields $f (x), X^A (x)$.
Variation of $f$ yields the equation of motion $\partial_{\alpha}(\sqrt{\eta}\eta^{\alpha \beta}\partial_{\beta}f)= 0$,
which is just the flat spacetime equation $\eta^{AB}\partial_A\partial_B f=0$ written in the coordinates $x^{\alpha}$.
On varying $X^A$, one obtains equations which are satisfied if  $\eta^{AB}\partial_A\partial_B f=0$.
This implies that $X^A(x)$ are   undetermined functions (subject to the condition that determinant of $\partial_{\alpha}X^A$
is non- vanishing). This 2 functions- worth of gauge is a reflection of the 2 dimensional diffeomorphism invariance
of $S_{PFT}$. Clearly the dynamical content of $S_{PFT}$ is the same as that of $S_0$; it is only that
the diffeomorphism invariance of
$S_{PFT}$ naturally allows a description of the standard free field dynamics dictated by $S_0$ on {\em arbitrary}
foliations of the fixed flat spacetime.

\subsection{Hamiltonian Formulation of PFT.}

In the previous subsection, $X^{A}(x)$ had a dual interpretation - one as  dynamical variables to be varied in the action, and
the other as inertial coordinates on a flat spacetime. In what follows we shall freely go between these two interpretations.

We set $x^0=t$ and $\{x^{\alpha}\} = \{t, x\}$. We restrict attention to $X^A(x)$ such that for any fixed
$t$, $X^A(t,x^a)$ describe an embedded spacelike hypersurface in the 2 dimensional flat spacetime (it is for this
reason that $X^A(x)$ are called embedding variables in the literature). This means that, for fixed $t$, the functions $X^A(x)$
must be such that the symmetric form $q_{ab}$ defined by
\begin{equation}
q_{ab}(x):=\eta_{AB}\frac{\partial X^A(x)}{\partial x}\frac{\partial X^B(x)}{\partial x}
\label{nondeg}
\end{equation}
is an 1 dimensional Riemannian metric. This follows from the fact that $q_{ab}(x)$ is the induced metric
on the hypersurface in the flat spacetime defined by $X^A(x)$ at fixed $t$.

A 1+1 decomposition of $S_{PFT}$ with respect to the time `$t$', leads to its Hamiltonian form:
\begin{equation}
S_{PFT} [f,X^A;\pi,\Pi_A;N^A]
=\int dt \int d^{n}x (\Pi_A{\dot X}^A + \pi_f {\dot f}- N^A H_A).
\label{spftham}
\end{equation}
 Here $\pi_f$ is the momentum conjugate to the scalar field $f$, $\Pi_A$ are the
momenta conjugate to the embedding variables $X^A$, $N^A$ are Lagrange multipliers for the first class
constraints $H_A$.
It turns out that the motions on phase space
generated by the `smeared'  constraints, $\int d^nx (N^AH_A)$ correspond to
scalar field evolution along arbitrary foliations of the flat spacetime, each choice of foliation being in correspondence with
a choice of multipliers $N^A$. Since the constraints are first class they also generate gauge transformations,
and as in General Relativity, the notions of gauge and evolution are intertwined.

Since free scalar field theory in 2 dimensions finds its simplest expression in terms of left and right movers,
it is useful to make a point canonical transformation to light cone embedding variables $X^{\pm}(x):= T(x)\pm X(x)$ (here we have set $X^0=T, X^1=X$). Denoting the conjugate embedding momenta by $\Pi^{\pm}(x)$, and setting
$H_{\pm}= H_0\pm H_1$,
the action takes the form
\begin{equation}\label{eq:1}
S\ =\ \int dt\int dx\ [\ \pi_{f}\dot{f}\ +
\Pi_{+}\dot{X^{+}}\ +\ \Pi_{-}\dot{X^{-}}\ -\
N^{+}{{H_{+}}}\ -\ N^{-}{{H_{-}}}\ ].
\end{equation}
where $N^{\pm}$ are the new Lagrange multipliers appropriate to $H_{\pm}$. Explicitly, the constraints $H_{\pm}$
are given by 
\begin{equation}\label{eq:2}
{{H_{\pm}}}(x)\ =\ [\ \Pi^{\pm}(x)X^{\pm'}(x)\ \pm\
\frac{1}{4}(\pi_{f}\ \pm f')(x)(\pi_{f}\ \pm f')(x)\ ].
\end{equation}
Note that while $X^{\pm}(x), f(x)$ transform as scalars under spatial coordinate transformations,
$\Pi_{\pm},\pi_f, N^{\pm}$ transform as scalar densities (or equivalently as spatial vector fields).

The Poisson brackets between various fields are given by,\\
\begin{equation}\label{eq:4}
\begin{array}{lll}
\lbrace f(x), \pi_{f}(x')\rbrace\ =\ \delta(x,x'),\\
\vspace*{0.1in}
\lbrace X^{\pm}(x), \Pi_{\pm}(x')\rbrace \ =\ \delta(x,x'),
\end{array}
\end{equation}
and the remaining brackets are zero.
Here $\delta(x,x')$ is the  delta-function on $S^{1}$.

To complete the transition to variables closely related to the left and right movers of free scalar field theory 
\cite{karelcm},
we
perform a canonical transformation on the matter variables. $(f, \pi_{f})\
\rightarrow (Y^{+}, Y^{-})$. Here  $Y^{\pm}(x)\ =\ \pi_{f}(x)\ \pm
f'(x)$ (strictly speaking this  transformation is not invertible when the spatial topology is $S^1$ due to the 
existence of zero modes; we shall return to this issue in section 3).
The Poisson brackets between  the scalar densities, $Y^{\pm}$, are given by,
\begin{equation}\label{eq:5}
\begin{array}{lll}
\lbrace Y^{\pm}(x), Y^{\pm}(x')\rbrace\ =\
\pm[\partial_{x}\delta(x,x')\ -\ \partial_{x'}\delta(x',x)]\\
\vspace*{0.1in}
\lbrace Y^{\mp}(x), Y^{\pm}(x')\rbrace\ =\ 0.
\end{array}
\end{equation}
The constraints are now
\begin{equation}\label{eq:2pm}
{{H^{\pm}}}(x)\ =\ [\ \Pi_{\pm}(x)X^{\pm'}(x)\ \pm\
\frac{1}{4}Y^{\pm}(x)^2\ ].
\end{equation}
and the constraint algebra is
\begin{equation}\label{eq:6}
\begin{array}{lll}
\lbrace {{H}}_{\pm}[N^{\pm}],{{H}}_{\pm}[M^{\pm}]
\rbrace\ =\ {{H}}_{\pm}[\mathcal{L}_{N^{\pm}}M^{\pm}]\\
\vspace*{0.1in}
\lbrace {{H}}_{\pm}[N^{\pm}],{{H}}_{\mp}[M^{\mp}]
\rbrace\ =\ 0
\end{array}
\end{equation}
Here
$\mathcal{L}_{N}$ denotes the Lie derivative with respect to the 1 dimensional spatial vector field
with component $N(x)$ in the coordinate system `x'.
The action of the constraints on the phase space variables can be expressed
as follows. Let $\Phi^{\pm}\ =\ (Y^{\pm}, \Pi_{\pm})$, we have
\begin{equation}\label{eq:7}
\begin{array}{lll}
\lbrace \Phi^{\pm}(x),\ {{H_{\pm}}}[N^{\pm}]\ \rbrace\ =\
\mathcal{L}_{N^{\pm}}\Phi^{\pm}(x)\\
\lbrace \Phi^{\mp}(x),\ {{H_{\pm}}}[N^{\pm}]\ \rbrace\ =\ 0,
\end{array}
\end{equation}
Thus, on the set of variables $\Phi^{\pm}$, infintesmal gauge transformations act as diffeomorphisms on $S^1$
and there is a split of the constraints and the phase space variables into commuting `+' and `-' parts which
correspond to the usual right and left moving sectors of free scalar field theory.
The action of the constraints on the embedding variables $X^{\pm}(x)$ preserves this split:
\begin{eqnarray}
\{ X^{\pm}(x),{{H_{\pm}}}[N^{\pm}]\} &=& N^{\pm}(X^{\pm})^{\prime},\label{c,x1}\\
\{ X^{\mp}(x),{{H_{\pm}}}[N^{\pm}]\} &=& 0.
\label{c,x2}
\end{eqnarray}
Indeed, the above equations seem to indicate that infinitesimal gauge transformations, once again, act as
diffeomorphisms on $S^1$; however, as we shall see in the next subsection, this interpretation is not strictly true for
equations (\ref{c,x1}), (\ref{c,x2})
due to
the non- existence of global, single valued coordinates on $S^1$.

\subsection{Consequences of spatial topology = $S^1$.}
\subsubsection{Conditions on the canonical variables.}
$S^1$ does not admit a global single valued coordinate system. However, at the cost of introducing appropriate
periodic/quasiperiodic boundary conditions on the fields we may choose $x$ to be the standard angular coordinate,
$x\in [0,2\pi]$ with the identification $x=0 \sim x=2\pi$. The Minkowskian coordinates $X^A= (T,X)$ in the action
(\ref{s0}) are chosen so that $T\in (-\infty, \infty)$,$X\in (-\infty, \infty)$ with the identifications
$X\sim X+2\pi$.

The above specifications on $x,X$ imply the following conditions on the canonical embedding variables and the Lagrange
multipliers:\\
\noindent{\bf (i)} $X^{\pm} (2\pi)- X^{\pm}(0) = 2\pi$.\\
\noindent{\bf (ii)} Any two sets of embedding data $(X_1^+(x), X_1^-(x))$
and $(X_2^+(x), X_2^-(x))$ are to be identified if there exists an integer $m$ such that
$X_{1}^{+}(x) = X_{2}^{+}(x) + 2m\pi\ \forall\ x\in\ [0,2\pi]$ and 
$X_{1}^{-}(x) = X_{2}^{-}(x) - 2m\pi\ \forall\ x\in\ [0,2\pi]$.\\

\noindent{\bf(iii)} $\Pi_{\pm}(x), N^{\pm}(x)$ and their spatial derivatives to all orders, as well
as the spatial derivatives to all orders of the embedding coordinates $X^{\pm}(x)$
are periodic on $[0,2\pi]$ with period $2\pi$.
This follows from the 1+1 Hamiltonian decomposition of (\ref{spft}) and
the fact that $\frac{\partial X^A}{\partial x^{\alpha}}$ in equation (\ref{nondeg})
is single valued on $S^1\times R$. \\
An additional ``non-degeneracy'' condition arises from (\ref{nondeg}):\\
\noindent{\bf (iv)}$\pm(X^{\pm})^{\prime} > 0$.

Since $f$ in (\ref{s0}) is a single valued function on $S^1\times R$, it follows that the matter phase space
variables, $(f,\pi_f)$ and their spatial derivatives to all orders are also periodic functions on $[0,2\pi]$.
Note also that the delta function $\delta (x,y)$ in (\ref{eq:4}), (\ref{eq:5}) is periodic in both its arguments.

\subsubsection{Finite gauge transformations.}

Whereas
equation (\ref{eq:7}) implies that finite gauge transformations act on $(\Pi_{\pm}, Y^{\pm})$ as spatial diffeomorphisms
on $S^1$, as remarked earlier the
case of the embedding variables $X^{\pm}$ is more subtle as $X^{\pm}$ are not single valued fields on $S^{1}$
by virtue of {\bf (i)}, section 2.3.1. Therefore,
evolution of $X^{\pm}$ under the flow generated by the constraints is better understood in terms of
transformations on the universal cover of $S^{1}$ as follows.

Unwind $S^1$ to its universal cover $\mathbf{R}$.
Quasi-periodic boundary conditions obeyed by the embeddings suggest that their extension to $\mathbf{R}$ satisfies:
\begin{equation}\label{eq:11}
X^{\pm}_{ext}(x\pm 2n\pi)\ :=\ X^{\pm}(x)\ \pm\ 2n\pi
\end{equation}
where $x\in [0,2\pi]$ and $n\in \mathbf{Z}$.
The vector fields $N^{\pm}(x)$ on $S^{1}$ extend to periodic vector fields  $N^{\pm}_{ext}$ on $\mathbf{R}$
so that $N^{\pm}_{ext}(x+2n\pi)\ =\ N^{\pm}(x),\;x\in[0,2\pi]$.
Let the 1 parameter family of (periodic) diffeomorphisms of $\mathbf{R}$ generated by $N^{\pm}_{ext}$ be denoted
by $\phi(N^{\pm}_{ext},t)$. Then it is straightforward to check that
the finite transformations generated by the constraints on $X^{\pm}(x)$ are labelled by
$\phi[N^{\pm}_{ext},t]$ and act as follows:
\begin{equation}\label{eq:13}
\begin{array}{lll}
(\alpha_{\phi[N^{\pm}_{ext},t]}X^{\pm})(x)\ =\ X^{\pm}_{ext}(\phi[N^{\pm}_{ext},t](x))\ \forall\ x\in [0,2\pi]\\
\vspace*{0.1in}
(\alpha_{\phi[N^{\pm}_{ext},t]}X^{\mp})(x)\ =\ X^{\mp}(x)\ \forall\ x\in [0,2\pi]
\end{array}
\end{equation}
Here
$\alpha_{\phi[N^{\pm}_{ext},t]}$ is the flow generated by Hamiltonian vector
field of ${{H_{\pm}}}[N^{\pm}]$.

It is also straightforward to see that the action of finite gauge transformations on 
the phase space variables $\Phi^{\pm}\in \{ Y^{\pm}, \Pi_{\pm}\}$ can equally well be written in terms of
the action of the periodic diffeomorphisms $\phi[N^{\pm}_{ext},t]$ on the periodic extensions $\Phi^{\pm}_{ext}$
as
\begin{equation}\label{eq:13**}
\begin{array}{lll}
(\alpha_{\phi[N^{\pm}_{ext},t]}\Phi^{\pm})(x)\ =\ \Phi^{\pm}_{ext}(\phi[N^{\pm}_{ext},t](x))\ \forall\ x\in[0,2\pi]\\
\vspace*{0.1in}
(\alpha_{\phi[N^{\pm}_{ext},t]}\Phi^{\mp})(x)\ =\ \Phi^{\mp}(x)\ \forall\ x\in[0,2\pi]
\end{array}
\end{equation}
Here $\Phi^{\pm}_{ext}(x+2n\pi)\ =\ \Phi^{\pm}(x)\;\forall x\in [0,2\pi], n\in \mathbf{Z}$.

Since $\phi[N^{\pm}_{ext},t], \;\forall (N^{\pm}_{ext},t)$ range over all periodic diffeomorphisms of
$\mathbf{R}$ connected to identity, we label every finite gauge transformation by a pair of such 
diffeomorphisms $(\phi^+,\phi^-)$ so that the Hamiltonian flows generated by $H_{\pm}$ are denoted by 
$\alpha_{\phi^{\pm}}$. 
To summarise: Let $\Psi^{\pm}(x)\in (X^{\pm}(x), \Pi_{\pm}(x), Y^{\pm}(x))$ 
and let its appropriate quasiperiodic/periodic  extension on $\mathbf{R}$ be $\Psi^{\pm}_{ext}$.
Then  we have that, $\forall x\in[0,2\pi]$,
\begin{equation}\label{eq:13*}
\begin{array}{lll}
(\alpha_{\phi^{\pm}}\Psi^{\pm})(x)\ =\ \Psi^{\pm}_{ext}(\phi^{\pm}(x))\\
\vspace*{0.1in}
(\alpha_{\phi^{\pm}}\Psi^{\mp})(x)\ =\ \Psi^{\mp}(x).
\end{array}
\end{equation}
Equations (\ref{eq:13*}) imply a left representation of the group of periodic diffeomorphisms of $\mathbf{R}$ by the
Hamiltonian flows corresponding to finite gauge transformations:
\begin{eqnarray}
\alpha_{\phi_1^{\pm}}\alpha_{\phi_2^{\pm}}&=& \alpha_{\phi^{\pm}\circ \phi_2^{\pm}} \label{alphaphi1phi2}\\
\alpha_{\phi_1^{\pm}}\alpha_{\phi_2^{\mp}}&=& \alpha_{\phi_2^{\mp}}\alpha_{\phi_1^{\pm}}. 
\end{eqnarray}

We emphasize that 
the  extended fields are only formal
constructs which are useful for interpreting gauge
transformations in terms periodic  diffeomorphisms of $\mathbf{R}$. The spatial slice
is always $S^1$ coordinatized by $x\in[0,2\pi]$ with boundary points
identified.

\subsection{Dirac Observables}

Since finite gauge transformations act as periodic diffeomorphisms of $\mathbf{R}$, it follows, directly,
that the integral over $x\in[0,2\pi]$ of any periodic scalar density constructed solely from the phase space
variables, is an observable. 

An analysis of the Hamiltonian equations \cite{karelcm} shows that the relation 
between solutions $f(X^+, X^-)$ of the flat spacetime wave equation and canonical data $(Y^{\pm}, X^{\pm})$ on
the constraint surface is
\begin{equation}
\pm 2\frac{\partial f}{\partial X^{\pm}}= \frac{Y^{\pm}}{(X^{\pm})^{\prime}}.
\label{fpmypm}
\end{equation}
Here $f$ is evaluated at the spacetime point $(X^+, X^-)$ defined by the canonical data.
Recall that any solution $f(X^+, X^-)$ to the free scalar field equation is of the form
\begin{equation}
f(X^+, X^-) = {\bf q} + {\bf p} \frac{(X^+ + X^-)}{2} -i
\sum_{n=1}^{\infty} ({\bf a_{(+)n}}e^{-inX^+} +{\bf a_{(-)n}}e^{-inX^+} +{\rm c.c}),
\label{fexp}
\end{equation}
where c.c. stands for `complex conjugate'. Equations (\ref{fpmypm}) and (\ref{fexp}) yield an interpretation
for the Dirac observables constructed below.

\subsubsection{Mode functions.}
From (\ref{fpmypm}) and (\ref{fexp}) and the remarks above, it follows that
\begin{equation}
a_{(\pm)n}= \int_{S^1}dx Y^{\pm}(x) e^{inX^{\pm}(x)}, \;n\in {\mathbf Z}, \;n>0
\label{eq:16}
\end{equation}
(and their complex conjugates, $a_{(\pm)n}^*$,) are Dirac observables which correspond to the
mode functions ${\bf a_{{(\pm)}n}}$ of equation (\ref{fexp}).
These observables form the (Poisson) algebra,
\begin{equation}\label{eq:17}
\begin{array}{lll}
\lbrace a_{n}, a_{m}* \rbrace\ =\ -4\pi in\delta_{n,m},\\
\lbrace a_{n}, a_{m} \rbrace\ =\ 0,\\
\lbrace a_{n}*, a_{m}* \rbrace\ =\ 0.
\end{array}
\end{equation}

\subsubsection{Zero modes.}
The quantities ${\bf q}, {\bf p}$ in equation (\ref{fexp}) are referred to as zero modes of the scalar field
and are also realizable as Dirac observables
which are canonically conjugate to each other \cite{karelcm}. Indeed, it is straightforward to see from 
(\ref{fpmypm}), (\ref{fexp}) that ${\bf p}$ corresponds to $p:= \int_{S^1}dxY^+(x)= \int_{S^1}dxY^-(x)$.
However, the degree of freedom corresponding to ${\bf q}$ is absent in the phase space coordinates 
$(X^{\pm}, \Pi_{\pm}, Y^{\pm})$ as a result of $Y^{\pm}$ only containing derivatives of $f$ (see equation
(\ref{fpmypm})).

Our aim in this work is to construct a triangulation independent polymer quantization of a generally covariant
field theoretic model. Issues related to the construction of zero modes (which are anyway mechanical (as opposed to 
field theoretic) degrees of freedom)  as Dirac observables serve to 
distract from this aim. Hence we shall switch off the zero modes by setting ${\bf q}={\bf p}=0$. Since 
${\bf q}$ and ${\bf p}$ are canonically conjugate, this can be done consistently.
In the free scalar field action (\ref{s0}) this corresponds to limiting the space of all scalar fields
by the conditions ${\bf q} =\int_{S^1} dX f(T,X) =0$ and ${\bf p} =\int_{S^1} dX \frac{\partial f(T,X)}{\partial T}=0$.
In the canonical description of PFT in terms of $(\Pi_{\pm}, X^{\pm}, Y^{\pm})$, since ${\bf q}$ does not appear, 
we only need to set the quantity 
\begin{equation}\label{eq:17a}
p:= \int_{S^1}dxY^+(x)= \int_{S^1}dxY^-(x)= 0.
\label{p=0}
\end{equation}
Since, as can easily be checked, $p$ commutes with $(\Pi_{\pm}, X^{\pm}, Y^{\pm})$ as well as the constraints
(\ref{eq:2pm}), it is consistent to impose (\ref{p=0}).

To summarize: The system we consider in this work is PFT on $S^1\times R$ with the zero modes switched off.
The phase space variables are 
$(\Pi_{\pm}, X^{\pm}, Y^{\pm})$ subject to the conditions of section 2.3.1. The symplectic structure is given by 
(\ref{eq:4}) and (\ref{eq:5}) and the constraints by (\ref{eq:2pm}). The degrees of freedom of the theory reside
entirely in the mode coefficients ${\bf a_{(\pm)n}},{\bf a_{(\pm)n}}^*$ (\ref{fexp}) which are expressed as the
functions $a_{(\pm)n},a^*_{(\pm)n}$ on phase space via (\ref{eq:16}).

\subsubsection{Conformal Isometries.}

Free scalar field theory in 1+1 dimensions (\ref{s0}) is conformally invariant. As a consequence  
the generators of conformal isometries in PFT are also Dirac observables (for details, see Reference \cite{karelcm}).
Consider the conformal isometry generated by the conformal Killing field ${\vec U}$ on the Minkowskian cylinder.
Let ${\vec U}$ have the components $(U^+(X^+), U^-(X^-))$ in the $(X^+, X^-)$ coordinate system.
$U^{\pm}$ are periodic functions of $X^{\pm}$ by virtue of the fact that ${\vec U}$ is smooth vector field on
the flat spacetime $S^1\times R$
These components of ${\vec U}$ naturally correspond to the functions 
 $(U^+(X^+(x)), U^-(X^-(x)))$ on the phase space of PFT. 
The Dirac observable corresponding to the generator of
conformal transformations associated with ${\vec U}$ is given by 
\begin{equation}\label{eq:18}
\Pi_{\pm}[U^{\pm}]\ =\ \int_{S^1} \Pi_{\pm}(x)U^{\pm}(X^{\pm}(x))
\end{equation}

These observables generate a Poisson algebra isomorphic to that of the commutator algebra of conformal Killing fields:
\begin{equation}\label{eq:19}
\begin{array}{lll}
\lbrace \Pi_{\pm}[U^{\pm}], \Pi_{\pm}[V^{\pm}]\rbrace\ =\
\Pi[[U,V]^{\pm}]\\
\vspace*{0.1in}
\lbrace \Pi_{\pm}[U^{\pm}], \Pi_{\mp}[V^{\mp}]\rbrace\ = 0.
\end{array}
\end{equation}
Here $[U,V]^{\pm}$ refer to the $\pm$ components of the commutator of the spacetime vector fields
${\vec U}, {\vec V}$. $[U,V]^{\pm}$ define functions of the embedding variables $X^{\pm}(x)$
in the manner described above.

Note that these observables are weakly equivalent, via the constraints (\ref{eq:2pm}) to
quadratic combinations of the mode functions \cite{karelcm}.
In the 
standard Fock representation of quantum theory (see for e.g. Reference \cite{karelqm}), these
quadratic combinations are nothing but the generators of the Virasoro algebra.

As we shall see, the polymer quantization of PFT provides a representation for the finite canonical transformations
generated by $\Pi^{\pm}[U^{\pm}]$. For future reference,
it is straightforward to check that the Hamiltonian flow, $\alpha_{(\Pi_{\pm}[U^{\pm}],t)}$ generated by
$\Pi_{\pm}[U^{\pm}]$ leaves the matter sector of phase space untouched and acts on the embedding variables
$X^{\pm}$ as
\begin{equation}
\alpha_{(\Pi_{\pm}[U^{\pm}],t)} X^{\pm}(x) = (\phi_{(\vec{U},t)} X^{\pm})(x).
\label{alphapiu}
\end{equation}
Here
$\phi_{(\vec{U},t)}$ denotes the one parameter family of conformal isometries generated by the 
conformal Killing field ${\vec U}$ on spacetime. $\phi_{(\vec{U},t)}$ maps the spacetime point $(X^+, X^-)$
to $\phi_{(\vec{U},t)} X^{\pm}$ and hence maps the  spatial slice 
defined by the canonical data $X^{\pm}(x)$ is mapped to the new slice (and hence the new canonical data)
$(\phi_{(\vec{U},t)} X^{\pm})(x)$.

$\phi_{(\vec{U},t)}$ ranges over all conformal isometries connected to identity. Any such conformal isometry
$\phi_c$ is specified by a pair of functions $\phi_c^{\pm}$ so that 
$\phi_c(X^+, X^-) := (\phi_c^+ (X^+), \phi_c^- (X^-))$. Invertibilty of $\phi_c$ together with connectedness with identity
 implies that 
\begin{equation}
\frac{d\phi_c^{\pm}}{dX^{\pm}} >0,
\label{phicnondeg}
\end{equation}
and the cylindrical topology of spacetime implies
that
\begin{equation}
\phi_c^{\pm}(X^{\pm}\pm 2\pi) = \phi_c^{\pm}(X^{\pm})\pm 2\pi.
\label{phicqp}
\end{equation}
Thus, we may denote the Hamiltonian flows which generate conformal isometries by 
$\alpha_{\phi_c}$ or, without loss of generality, by $\alpha_{\phi_c^{\pm}}$ with
$\alpha_{\phi_c^{\pm}}$ acting trivially on the $\mp$ sector.

To summarise:$\alpha_{\phi_c^{\pm}}$ leave the matter variables untouched, so that
\begin{equation}
\alpha_{\phi_c^{\pm}} Y^{\pm}(x) = Y^{\pm}(x),\;\;\;
\alpha_{\phi_c^{\pm}} Y^{\mp}(x)=Y^{\mp}(x),
\label{phicalpha0}
\end{equation}
 and act on $X^{\pm}(x)$ as
\begin{equation}
\alpha_{\phi_c^{\pm}} X^{\pm}(x) = \phi_c^{\pm} (X^{\pm}(x)),\;\;\;
\alpha_{\phi_c^{\pm}} X^{\mp}(x)=X^{\mp}(x).
\label{phicalpha1}
\end{equation}
Further, since $\Pi_{\pm}[U^{\pm}]$ are observables which commute strongly with the constraints,
the corresponding Hamiltonian flows are gauge invariant. This translates to the condition that for all
\begin{equation}
\begin{array}{lll}
\alpha_{\phi_c^{\pm}}\circ\alpha_{\phi^+}\ =\ \alpha_{\phi^+}\circ\alpha_{\phi_c^{\pm}}\\
\vspace*{0.1in}
\alpha_{\phi_c^{\pm}}\circ\alpha_{\phi^-}\ =\ \alpha_{\phi^-}\circ\alpha_{\phi_c^{\pm}}
\label{phicalpha2}
\end{array}
\end{equation}
where as before $\phi^{\pm}$ label finite gauge transformations.

\section{Polymer Quantum Kinematics.}
\subsection{Preliminaries.}
As in LQG, the polymer quantization is based on suitably defined ``holonomies'' and the polymer Hilbert
space is spanned by suitably defined ``charge network'' states. In view of the correspondence between 
finite gauge transformations and periodic diffeomorphisms of $\mathbf{R}$, it is useful to 
to define periodic and quasiperiodic extensions of charge network labels. Hence we define the following.\\

\noindent
{\bf Definition 1} :  
A charge-network  $s$ is specified by  the  labels
$(\gamma(s),(j_{e_{1}},...,j_{e_{n}}))$ consisting of a graph $\gamma (s)$ (by which we
mean a finite collection of closed, non-overlapping(except in boundary
points) intervals which cover $[0,2\pi]$)  and `charges'  $j_{e}\in \mathbf{R}$ assigned to each
interval e. (Note that $j_{e}=0$ is allowed.)
Equivalence classes of charge- networks are defined as follows.
The charge- network $s^{\prime}$ is said to be finer than $s$ iff (a) every edge of $\gamma (s)$ is identical to,
or composed of, edges in $\gamma (s^{\prime})$ (b) the charge labels of identical edges in 
$\gamma (s),\gamma (s^{\prime})$
are identical and the charge labels of the edges of $\gamma (s^{\prime})$ which compose to yield an edge of 
$\gamma (s)$ are identical and equal to that of their union in $\gamma (s)$.
Two charge- networks are equivalent  if there exists a charge- network finer than both. 
Hence we can represent each equivalence class by a unique representative $s$
such that no two adjacent edges have the same charge. However, unless otherwise mentioned, $s$ will not 
necessarily denote this unique choice.

\noindent
{\bf Definition 2}: The periodic extension of the charge- network $s$ to $\mathbf{R}$ is denoted by $s_{ext}$ and
 defined as follows.

Given a graph $\gamma$ as in Definition 1 above, $T_{N}(\gamma)$ denotes the translation of $\gamma$ by $2N\pi$, 
i.e. $T_{N}(\gamma)$ lies in $[2N\pi, 2(N+1)\pi]$. We define the {\em extension} of $\gamma$ to $\mathbf{R}$ as  
$\gamma_{ext}\ =\ \cup_{N\in\mathbf{Z}}\ T_{N}(\gamma)$.
The {\em restriction} of $\gamma_{ext}$ to any interval $I\subset \mathbf{R}$ is denoted by
$\gamma_{ext}\vert_I$ so that $\gamma_{ext}\vert_{[0,2\pi]}= \gamma$.

Given a charge network $s=(\gamma(s),(j_{e_{1}},...,j_{e_{n}}))$, $s_{ext}$ is specified by the graph 
$\gamma (s_{ext}):=\gamma (s)_{ext}$ ($\gamma (s)_{ext}$ denotes the extension of $\gamma (s)$ to $\mathbf{R}$)
and charge labels for each edge of $\gamma (s_{ext})$ which are such that 
$T_{N}(\gamma (s))\subset \gamma (s_{ext})$ has  the same set of charges which are on $\gamma$. Thus\\
\hspace*{0.4in} 1. On any closed interval $I_{N}\ =\ [2N\pi,\ 2(N+1)\pi]$, $N\in\mathbf{Z}$,
$\gamma (s_{ext})\vert_{I_{N}}\cong \gamma(s)$.\\
\hspace*{0.4in} 2. The set of charges on $\gamma(s_{ext})\vert_{I_{N}}$ is $(j_{e_{1}},...,j_{e_{n}})$.\\
\noindent
We refer to $s_{ext}\vert_{[0,2\pi]}$ as the restriction of $s_{ext}$ to $[0,2\pi]$ so that 
$s_{ext}\vert_{[0,2\pi]}\ =\ s$.\\

\noindent
{\bf Definition 3}: The quasi- periodic 
extension of the charge- network $s$ to $\mathbf{R}$ is  denoted by ${\bar s}_{ext}$ and defined as follows.
Given a charge network $s=(\gamma(s),(j_{e_{1}},...,j_{e_{n}}))$, 
${\bar s}_{ext}$ is specified by the graph 
$\gamma ({\bar s}_{ext}):=\gamma (s)_{ext}$ and charge labels for each edge of $\gamma ({\bar s}_{ext})$ 
which are such that 
$T_{N}(\gamma (s))\subset \gamma ({\bar s}_{ext})$ has  the  set of charges which are on $\gamma$ augmented by $2N\pi$.
Thus\\
\hspace*{0.4in} 1. On any closed interval $I_{N}\ =\ [2N\pi,\ 2(N+1)\pi]$, $N\in\mathbf{Z}$,
$\gamma({\bar s}_{ext})\vert_{I_{N}}\cong \gamma(s)$.\\
\hspace*{0.4in} 2. The set of charges on $\gamma ({\bar s}_{ext})\vert_{I_{N}}$ is $(j_{e_{1}}\ +\ 2N\pi,...,j_{e_{n}}\ +\ 2N\pi)$.\\

\noindent {\bf Definition 4}: The action of periodic diffeomorphisms on $\gamma_{ext}, s_{ext}, {\bar s}_{ext}$
may be defined as follows.
Any periodic diffeomorphism $\phi$ of $\mathbf{R}$ commutes with the $2\pi$ translations, $T_N$ and hence
has a natural action on the extension $\gamma_{ext}$ of the graph $\gamma$. Denote the resulting graph by 
$\phi (\gamma_{ext})$ and let the edge $\phi (e)\in \phi (\gamma_{ext})$ be the image, by $\phi$ of the edge
$e \in \gamma_{ext}$. The action of $\phi$ on the extensions $s_{ext}, {\bar s}_{ext}$ is defined by \\
\noindent (i) mapping the underlying graph $\gamma (s)_{ext}$ to $\phi (\gamma (s)_{ext})$\\
\noindent (ii) labelling the edge $\phi (e) \in \phi (\gamma (s)_{ext})$ by the same charge as 
the edge $e \in \gamma (s)_{ext}$ so that $k_{\phi (e)}= k_e$.

Denote the resulting periodic/quasiperiodic charge networks on $\mathbf{R}$ by $\phi(s_{ext})$/$\phi({\bar s}_{ext})$

\subsection{Embedding sector.}
\subsubsection{The *- Algebra}

The elementary variables which generate  the *-Poisson algebra 
are,  $X^{+}(x),T_{s^+}[\Pi_{+}]$,
 $X^{-}(x),
T_{s^-}[\Pi_{-}]$.  Here $T_{s^{\pm}}[\Pi_{{\pm}}]$ are the holonomy- type functions associated with 
the charge networks $s^{\pm}$ given by 
\begin{equation}\label{eq:20}
T_{s^{\pm}}[\Pi_{\pm}]\ =\ \prod_{e^{\pm}\in \gamma(s^{\pm})} \exp [-ik_{e^{\pm}}^{\pm} \int_{e^{\pm}}\Pi_{\pm}].
\end{equation}
The only non- trivial Poisson brackets are:
\begin{equation}\label{eq:21}
\begin{array}{lll}
\lbrace X^{\pm}(x), T_{s^{\pm}}[\Pi_{\pm}]\rbrace & = &
-ik_{e^{\pm}}^{\pm}T_{s^{\pm}}[\Pi_{\pm}]\ if\ x\in \mbox{Interior}(e^{\pm})\\
\vspace*{0.1in}
& = & -\frac{i}{2}(k_{e_{I^{\pm}}^{\pm}}^{\pm}+k_{e_{(I+1)^{\pm}}^{\pm}}^{\pm})T_{s^{\pm}}^{E}[\Pi_{\pm}]\ if\ x\in e_{I^{\pm}}^{\pm}\cap e_{(I+1)^{\pm}}^{\pm}\ 1\leq I^{\pm}\leq (n-1)^{\pm}\\
\vspace*{0.1in}
\lbrace X^{\pm}(0),T_{s^{\pm}}[\Pi_{\pm}]\rbrace & = & \lbrace
X^{\pm}(2\pi),T_{s^{\pm}}[\Pi_{\pm}]\rbrace\ =\
-\frac{i}{2}(k_{e_{1}^{\pm}}^{\pm}+k_{e_{n^{\pm}}^{\pm}}^{\pm})T_{s^{\pm}}[\Pi_{\pm}],
\end{array}
\end{equation}
where the last Poisson bracket uses the periodicity of delta
function.  The *-relations are given by
\begin{equation}\label{eq:22}
\begin{array}{lll}
(X^{\pm}(x))^{*}\ =\ X^{\pm}(x)\ \forall\ x\in\ [0,2\pi]\\
\vspace*{0.1in}
T_{s^{\pm}}[\Pi_{\pm}]^{*}\ =\ T_{-s^{\pm}}[\Pi_{\pm}], \; -s^{\pm}=(\gamma(s^{\pm}),(-k_{e_{1}^{\pm}}^{\pm},...,-k_{e_{n^{\pm}}^{\pm}}^{\pm}))
\end{array}
\end{equation}
The action of finite gauge transformations on these elementary functions is as follows
(we only analyse the right-moving sector; the analysis of the left moving sector is identical).

From equation (\ref{eq:13*}) we have,
\begin{equation}\label{eq:23*}
\alpha_{\phi^+}T_{s^{+}}[\Pi_{+}]\ =\ T_{s^{+}}[(\phi^+)_{*}\Pi_{+}].
\end{equation}
It is straightforward to check, using the periodicity of $\phi^+, \Pi_+ ,s^+_{ext}$  and the various definitions in
section 3.1 that
\begin{equation}
T_{s^{+}}[(\phi^+)_{*}\Pi_{+}] = T_{\phi^+(s^+_{ext})\vert_{[0, 2\pi]}} [\Pi_+ ].
\label{phits}
\end{equation}
Finite gauge transformations act on $X^{\pm}$ as in equations (\ref{eq:13}), (\ref{eq:13*}).
To summarise, under finite gauge transformations the generators of the Poisson algebra transform as:
\begin{equation}\label{eq:29}
\begin{array}{lll}
\alpha_{\phi^{\pm}}(X^{\pm})(x)\ =\ X^{\pm}_{ext}((\phi^{\pm})(x))\ =\ X^{\pm}(y)\ \pm 2\pi\ N\ \textrm{if}\ (\phi^{\pm})(x)=y+2\pi N\ y\in\ [0,2\pi]\\
\vspace*{0.1in}
\alpha_{\phi^{\mp}}(X^{\pm})(x)\ =\ X^{\pm}(x)\\
\vspace*{0.1in}
\alpha_{\phi^{\pm}}(T_{s^{\pm}}[\Pi^{\pm}])\ =\ T^{\pm}_{\phi^{\pm}(s^{\pm}_{ext})\vert_{[0,2\pi]}}[\Pi^{\pm}]\\
\vspace*{0.1in}
\alpha_{\phi^{\mp}}(T_{s^{\pm}}[\Pi^{\pm}])\ =\ T_{s^{\pm}}[\Pi^{\pm}]
\end{array}
\end{equation}

\subsubsection{Representation of the *- Algebra}
Denote the  kinematic Hilbert space for the $\pm$ embedding sectors by ${\cal H}^{\pm}_E$.
${\cal H}^{\pm}_E$ is the closure of the span of the orthonormal basis of  embedding `charge network states'.
Each such state is labelled by a charge network $s^{\pm}$ and denoted by $T_{s^{\pm}}$.
\footnote{More precisely, the labelling is by the equivalence class of $s^{\pm}$ as in Definition 1, section 3.1
\label{footnote2}}
The inner product is
\begin{equation}
< {T}_{s^{\pm}},\ {T}_{s^{\prime \pm}}>\ =\ \delta_{s^{\pm},s^{\prime \pm}}
\label{eip}
\end{equation}
where $\delta_{s^{\pm},s^{\prime \pm}}$ is a Kronecker delta function which is unity when the two charge networks
are identical and vanishes otherwise.

The `$\pm$' sector operators corresponding to the elementary functions of the previous section are denoted by
${\hat X}^{\pm} (x), {\hat T}_{s^{\pm}}$. ${\hat T}_{s^{\pm}}$ acts on the charge network states as:
\begin{equation}\label{eq:32}
\hat{T}_{s^{\pm}}T_{s^{\prime \pm}}\ :=\ T_{s^{\pm}+s^{\prime \pm}}
\end{equation}
where $s^{\pm}+s^{\prime \pm}$ is the charge network obtained by dividing $\gamma(s^{\pm})$, $\gamma(s^{\prime \pm})$
into maximal, non-overlapping (upto boundary points) intervals and assigning charge $k^{\pm}_{e^{\pm}}\ +\ k^{\pm}_{e'^{\pm}}$ to
$e^{\pm}\cap e'^{\pm}$ where $e^{\pm}\in \gamma(s^{\pm})$, $e'^{\pm}\in \gamma(s^{\pm}_{1})$.

The action of ${\hat X}^{\pm} (x)$ is:
\begin{equation}\label{eq:33}
\hat{X}^{\pm}(x)T_{s^{\pm}}\ :=\ \lambda_{x,s^{\pm}}T_{s^{\pm}},
\end{equation}
where, for $\gamma (s^{\pm})$ with $n^{\pm}$ edges,
\begin{equation}\label{eq:33a}
\begin{array}{lll}
\lambda_{x,s^{\pm}}:=\ \hbar k_{e_{I^{\pm}}^{\pm}}^{\pm}T_{s^{\pm}}\ if\ x\in \mbox{Interior}(e_{I^{\pm}}^{\pm})\ 1\leq I^{\pm}\leq n^{\pm}\\
\vspace*{0.1in}
\hspace*{0.3in}:=\ \frac{\hbar}{2} (k_{e_{I^{\pm}}^{\pm}}^{\pm}\
+\ k_{e_{(I+1)^{\pm}}^{\pm}}^{\pm})T_{s^{\pm}}\ if\ x\in e_{I^{\pm}}^{\pm}\cap e_{(I+1)^{\pm}}^{\pm}\ 1\leq I^{\pm}\leq (n-1)^{\pm}\\
\end{array}
\end{equation}
\begin{equation}\label{eq:33b}
\begin{array}{lll}
\hspace*{0.3in}:=\ \frac{\hbar}{2} (k_{e_{n^{\pm}}^{\pm}}^{\pm}\ \mp\ \frac{2\pi}{\hbar}\ +\ k_{e_{1}^{\pm}}^{\pm})T_{s^{\pm}}\ if\ x=0\\
\vspace*{0.2in}
\hspace*{0.3in}:=\ \frac{\hbar}{2} (k_{e_{1}^{\pm}}^{\pm}\ \pm\ \frac{2\pi}{\hbar}\ +\ k_{e_{n^{\pm}}^{\pm}}^{\pm})T_{s^{\pm}}\ if\ x=2\pi
\end{array}
\end{equation}
 The last two equations, (\ref{eq:33b}), implement the boundary condition $X^{\pm}(2\pi)-X^{\pm}(0)=\pm2\pi$ (see {\bf (i)} of
section 2.3.1.

It is straightforward to check that equations (\ref{eq:32}),(\ref{eq:33}),(\ref{eq:33a}),(\ref{eq:33b})
provide a representation of the Poisson bracket algebra (\ref{eq:21}) so that quantum commutators equal
$i \hbar$ times the Poisson brackets. It is also straightforward to verify that the *- relations (\ref{eq:22})
on ${\hat X}^\pm (x), {\hat T}_{s^\pm}$
are implemented by the inner product (\ref{eip}) so that ${\hat X}^\pm (x)$ are self adjoint and
${\hat T}_{s^\pm}$ are unitary.

\subsubsection{Unitary representation of finite gauge transformations.}
Since the Hamiltonian flows of $\alpha_{\phi^{\pm}}$ (\ref{eq:13*}) are real, the corresponding
quantum operators ${\hat U}(\phi^{\pm})$ must be unitary. Equations (\ref{eq:13*}), (\ref{alphaphi1phi2})
imply that this unitary representation must satisfy
\begin{equation}\label{eq:53}
\begin{array}{lll}
\hat{U}^{\pm}(\phi_{1}^{\pm})\hat{U}^{\pm}(\phi_{2}^{\pm})\ =\ \hat{U}^{\pm}(\phi_{1}^{\pm}\circ \phi_{2}^{\pm})\\
\vspace*{0.1in}
\hat{U}^{\pm}(\phi^{\pm})\hat{X}^{\pm}(x)\hat{U}^{\pm}(\phi^{\pm})^{-1}\
=\ \hat{X}^{\pm}(y^{\pm})\ \pm\ 2\pi N^{\pm}\\
\vspace*{0.1in}
\hat{U}^{\pm}(\phi^{\pm})\hat{T}_{s^{\pm}}\hat{U}^{\pm}(\phi^{\pm})^{-1}\
=\ \hat{T}_{\phi^{\pm}(s^{\pm})_{ext}\vert [0,2\pi]} .
\end{array}
\end{equation}
where $\phi^{\pm}(x)\ =\ y^{\pm} + 2\pi N^{\pm}$, with $y^{\pm}\in [0,2\pi]$ and $N^{\pm}\in \mathbf{Z}$.\\
We define the action of ${\hat U}(\phi^{\pm})$ to be
\begin{equation}\label{eq:56}
\begin{array}{lll}
\hat{U}^{\pm}(\phi^{\pm})T_{s^{\pm}}\ :=\ T_{\phi(\overline{s}^{\pm}_{ext})\vert_{[0,2\pi]}}\\
\vspace*{0.1in}
\hat{U}^{\mp}(\phi^{\mp})T_{s^{\pm}}\ :=\ T_{s^{\pm}}.
\end{array}
\end{equation}
The appearance of the quasi-periodic extensions ${\bar s}^{\pm}_{ext}$ of the charge networks $s^{\pm}$
(see Definition 3, section 3.1) in the first equation above may be anticipated from the quasi- periodic
nature of the embedding variables $X^{\pm}(x)$ (\ref{eq:11}). Unitarity of ${\hat U}^{\pm}(\phi^{\pm})$
follows straightforwardly:
\begin{equation}\label{eq:63}
\begin{array}{lll}
<\ \hat{U}^{\pm}(\phi^{\pm})T_{s_{1}^{\pm}}\ ,\ \hat{U}^{\pm}(\phi^{\pm})T_{s_{2}^{\pm}}\ > &
= & <\ T_{\phi(\overline{s}_{1}^{ext \pm}\vert_{[0,2\pi]})}\ ,\ T_{\phi(\overline{s}_{2}^{ext \pm}\vert_{[0,2\pi]})}\ >\\
\vspace*{0.1in}
& = & \delta_{\phi^{\pm}(\overline{s}_{1}^{ext \pm}\vert_{[0,2\pi]}),\ \phi^{\pm}(\overline{s}_{2}^{ext \pm}\vert_{[0,2\pi]})}\ \forall\ \phi^{\pm}\\
\vspace*{0.1in}
& = & \delta_{s_{1}^{\pm},s_{2}^{\pm}}
\end{array}
\end{equation}
where we have used the fact that two charge-networks are equal on $[0,2\pi]$ iff their extensions are equal.

From equation (\ref{eq:56}) and Definitions 3,4 of section 3.1, it follows that
\begin{equation}\label{eq:62}
\begin{array}{lll}
\hat{U}^{\pm}(\phi^{\pm}_{1})\hat{U}^{\pm}(\phi^{\pm}_{2})\ T_{s^{\pm}} & = & T_{\phi^{\pm}_{1}(\overline{\phi^{\pm}_{2}
(\overline{s^{\pm}}_{ext})\vert_{[0,2\pi]}})_{ext}\vert_{[0,2\pi]}}\\
\vspace*{0.1in}
& = & T_{\phi^{\pm}_{1}(\phi^{\pm}_{2}(\overline{s}^{\pm}_{ext}))\vert_{[0,2\pi]}}\\
\vspace*{0.1in}
& = & T_{(\phi^{\pm}_{1}\circ\phi^{\pm}_{2})(\overline{s}^{\pm}_{ext})\vert_{[0,2\pi]}}\\
\vspace*{0.1in}
& = & \hat{U}^{\pm}(\phi^{\pm}_{1}\circ\phi^{\pm}_{2})T_{s^{\pm}},
\end{array}
\end{equation}
thus verifying the first relation in (\ref{eq:53}).

Next, we turn to the second relation of (\ref{eq:53}). We sketch the proof for the `+' sector; the proof for the `-'
sector is on similar lines. From (\ref{eq:56}) and (\ref{eq:33}) we have that:
\begin{equation}\label{eq:64a}
\begin{array}{lll}
\hat{U}^{+}(\phi^{+})\hat{X}^{+}(x)\hat{U}^{+}(\phi^{+})^{-1}\ T_{s^{+}}\ &
=& \hat{U}^{+}(\phi^{+})\hat{X}^{+}(x)T_{(\phi^{+})^{-1}(\overline{s}_{ext}^{+}\vert_{[0,2\pi]}}\\
\vspace*{0.1in}
&=& \lambda_{x,(\phi^{+})^{-1}(\overline{s}_{ext}^{+}\vert_{[0,2\pi]})}T_{s^{+}}.
\end{array}
\end{equation}
It is straightforward to see that
\begin{equation}
\lambda_{x,(\phi^{+})^{-1}(\overline{s}_{ext}^{+}\vert_{[0,2\pi]})}\ =\ \lambda_{y^{+},s^{+}}+2\pi N^{+} ,
\end{equation}
which via equation (\ref{eq:33}) obtains the desired result.

Finally, we turn to the last relation of (\ref{eq:53}). Once again, we sketch the proof for the `+' sector; the
`-' sector proof follows analogously. We want to show that 
\begin{equation}\label{eq:66}
\hat{U}^{+}(\phi^+)\hat{T}_{s^{+}}\hat{U}^{+}((\phi^{+})^{-1})\ =\ \hat{T}_{\phi^+(s^{+}_{ext})\vert_{[0,2\pi]}}.
\end{equation}
Since charge network states form an orthonormal basis in the Hilbert space, it follows that (\ref{eq:66})
is equivalent to the condition that $\forall s^+_1, s^+_2$
\begin{equation}\label{eq:66a}
\langle T_{(\phi^+)^{-1}(\overline{s^+}_{1})_{ext}\vert_{[0,2\pi]}}\vert\ 
\hat{T}_{s^+}\vert\ T_{(\phi^+)^{-1}(\overline{s^+}_{2})_{ext}\vert_{[0,2\pi]}}\rangle 
= \langle T_{s^+_{1}}\vert\ \hat{T}_{\phi^+(s^+_{ext})\vert_{[0,2\pi]}}\vert T_{s^+_{2}}\rangle ,
\end{equation}
which from equation (\ref{eq:32}) is, in turn, equivalent to the equation
\begin{equation}\label{eq:66b}
\delta_{(\phi^+)^{-1}(\overline{s^+}_{1})_{ext}\vert_{[0,2\pi]},s^+ +
(\phi^+)^{-1}(\overline{s^+}_{2})_{ext}\vert_{[0,2\pi]}}\ =\ \delta_{s^+_{1},\phi^+(s^+_{ext})\vert_{[0,2\pi]}+s^+_{2}}.
\end{equation}
However, (suppressing the `+' superscript), we have that  
\begin{equation}\label{eq:66c}
\begin{array}{lll}
\delta_{\phi^{-1}(\overline{s}_{1})_{ext}\vert_{[0,2\pi]},s+\phi^{-1}(\overline{s}_{2})_{ext}\vert_{[0,2\pi]}} & = & \delta_{\phi^{-1}(\overline{s}_{1})_{ext},s_{ext}+\phi^{-1}(\overline{s}_{2})_{ext}}\\
& = & \delta_{(\overline{s}_{1})_{ext}, \phi(s_{ext})+ (\overline{s}_{2})_{ext}}\\
& = & \delta_{(s_{1})_{ext}, \phi(s_{ext})+ (s_{2})_{ext}}\\
& = & \delta_{s_{1}, \phi(s_{ext})\vert_{[0,2\pi]}+ s_{2}},
\end{array}
\end{equation}
thus proving (\ref{eq:66}).

\subsection{Matter sector.}
\subsubsection{The *- Algebra.}
The *- Algebra is generated by the operators corresponding to the classical holonomies
$W_{s^{\pm}}[Y^{\pm}]$ which are defined as
\begin{equation}\label{eq:38}
W_{s^{\pm}}[Y^{\pm}]\ =\ \exp[i \sum_{e^{\pm}\in E(\gamma(s^{\pm}))} l_{e^{\pm}}^{\pm} \int_{e^{\pm}}Y^{\pm}].
\end{equation}
Here $s^{\pm}\ :=\ \{ \gamma(s^{\pm}),\  (l_{e_{1}^{\pm}}^{\pm},...,l_{e_{m^{\pm}}^{\pm}}^{\pm})\ \}$ are charge- networks.
The algebra for the holonomy operators is the analog of the Weyl algebra for linear quantum fields.
Similar to that case, we need to first evaluate the Poisson brackets,
$\{ \sum_{e^{\pm}}l_{e^{\pm}}^{\pm}\int_{e^{\pm}}Y^{\pm},\ \sum_{e^{\prime\pm}}l_{e^{\prime\pm}}^{\pm}\int_{e^{\prime\pm}}Y^{\pm}\} \ $, between the exponents of  pairs of
classical holonomies and then use the Baker- Campbell- Hausdorff  Lemma \cite{bakercampbellhausdorff} to define
the algebra on the holonomy operators in quantum theory.

Let $\kappa_{e}$ be the characteristic function associated with a closed interval $e$ and denote the beginning and
final points of $e$ by $b(e)$ and $f(e)$ so that
\begin{equation}\label{eq:39}
\begin{array}{lll}
\kappa_{e}(x)\ =\ 1\ if\ x\in \mbox{Interior}(e)\\
\vspace*{0.1in}
\hspace*{0.5in}=\ \frac{1}{2}\ if\ x=b(e)\ or\ f(e)\\
\end{array}
\end{equation}
\begin{equation}\label{eq:39a}
\begin{array}{lll}
\hspace*{0.5in}=\ \frac{1}{2}\ if\ x=0\ $and$\ f(e)=2\pi\\
\vspace*{0.1in}
\hspace*{0.5in}=\ \frac{1}{2}\ if\ x=2\pi $and$\ b(e)=0.
\end{array}
\end{equation}
Here, equations (\ref{eq:39a}) follow from the  periodicity of the delta function. From equation (\ref{eq:5}) it follows
that
\begin{equation}\label{eq:40}
\{ \int_{e^{\pm}}Y^{\pm},\ \int_{e^{\prime\pm}}Y^{\pm}\}\ =\ \pm\alpha(e^{\pm},e^{\prime\pm})\
:=\ \pm(\kappa_{e^{\prime\pm}}\vert \partial_{e^{\pm}}\ -\ \kappa_{e^{\pm}}\vert \partial_{e^{\prime\pm}}),
\end{equation}
where
\begin{equation}
\kappa_{e}\vert \partial_{e'}:= \kappa_{e}(f(e'))\ -\ \kappa_{e}(b(e')),
\label{defkappavert}
\end{equation}
so that
\begin{equation}\label{eq:41}
\{ \sum_{e^{\pm}}l_{e^{\pm}}^{\pm}\int_{e^{\pm}}Y^{\pm},\ \sum_{e^{\prime\pm}}l_{e^{\prime\pm}}^{\pm}\int_{e^{\prime\pm}}Y^{\pm}\}\
=\ \pm\sum_{e^{\pm},e^{\prime\pm}}l_{e^{\pm}}^{\pm}l_{e^{\prime\pm}}^{\pm}\alpha(e^{\pm},e^{\prime\pm}).
\end{equation}
It follows that the `Weyl algebra' of holonomy operators is:
\begin{equation}\label{eq:42}
\begin{array}{lll}
\hat{W}(s^{\pm})\hat{W}(s^{\prime \pm})\ =\ \exp[\mp \frac{i\hbar}{2} \alpha(s^{\pm},s^{\prime\pm})]
\hat{W}(s^{\pm}+s^{\prime\pm}),\\
\vspace*{0.1in}
\hat{W}(s^{\pm})^{*}\ =\ \hat{W}(-s^{\pm}),
\end{array}
\end{equation}
where
\begin{equation}
\alpha(s^{\pm},s^{\prime\pm})\ :=\ \sum_{e^{\pm}\in \gamma(s^{\pm})}\sum_{e^{\prime\pm}\in \gamma(s^{\prime\pm})}l_{e}^{\pm}l_{e^{\prime\pm}}^{\pm}\alpha(e^{\pm},e^{\prime\pm}),
\label{defalphass'}
\end{equation}
with $\alpha(e,e')$ defined through equations (\ref{defkappavert}) and (\ref{eq:40}).
From the second equation of (\ref{eq:5}), it follows that the `+' and `-' holonomy operators commute, so that,
once again, these sectors can be treated independently.

\subsubsection{Representation of the *- Algebra.}
It is convenient to define the quantum theory through the Gelfand- Naimark - Segal (GNS) 
construction \cite{gns}. The explicit operator
action on the basis of charge network states is provided after we present the GNS state.

We define the GNS states $\omega_{\pm}$ on the $\pm$ holonomy algebras by specifying their action on the
holonomy operators as follows:
\begin{equation}\label{eq:47*}
\omega_{M}^{\pm}(\hat{W}(s^{\pm}))\ =\ \delta_{s^{\pm},\circ}.
\end{equation}
Here `$\circ$' is the trivial charge network which may be represented by graph $\gamma (\circ )$ consisting of the
single edge $e=[0, 2\pi ]$ with vanishing charge $l_e^{\pm}=0$. The Kronecker delta function
$\delta_{s^{\pm},\circ}$ is unity iff $s^{\pm}= \circ$ and vanishes otherwise.
It follows from the GNS construction that
the corresponding GNS Hilbert spaces ${\cal H}^{\pm}_M$ are spanned by charge network states denoted
by $W_{s^{\pm}}$. The inner product is
\begin{equation}\label{eq:48}
<\ W(s^{\pm}), W(s^{\prime\pm})\ >_{\pm}\ =\ \delta_{s^{\pm},s^{\prime\pm}}
\end{equation}
and the action of the holonomy operators is
\begin{equation}\label{eq:49}
\hat{W}(s^{\pm})W(s^{\prime\pm})\ =\ \exp[\mp \frac{i\hbar\alpha(s^{\pm},s^{\prime\pm})}{2}]W(s^{\pm}+s^{\prime\pm}).
\end{equation}
Here,  as for the embedding sector,
$s^{\pm}+s^{\prime\pm}$ is obtained by sub-dividing
$s^{\pm}$ and $s^{\prime\pm}$ into maximal non-overlapping (upto boundary points) intervals and putting
charges $l^{\pm}_{e}\ +\ l^{\pm}_{e^{\prime\pm}}$ on $e^{\pm}\cap e^{\prime\pm}$. ($e^{\pm}\in\ s^{\pm}$, $e^{\prime\pm}\in\ s^{\prime\pm}$).
\footnote{While our notation uses charge network labels, the operators ${\hat W}(s^{\pm})$ and states
$W(s^{\pm})$ only depend on the equivalence classes of labels. See also Footnote \ref{footnote2} in this regard.}

It is straightforward to check, explicitly, that equation (\ref{eq:49}) provides a representation for the first
equation of (\ref{eq:42}). Verification of the second equation of (\ref{eq:42}) is equivalent to showing that
$\forall s^{\pm},s^{\prime\pm},s^{\prime\prime\pm}$,
\begin{equation}
<\ W(s^{\prime\pm}),\ (\hat{W}(s^{\pm}))^{\dagger}W(s^{\prime\prime\pm})\ >_{\pm}
= <\ W(s^{\prime\pm}),\ \hat{W}(-s^{\pm})W(s^{\prime\prime\pm})>_{\pm}.
\label{wdagger}
\end{equation}
Equation (\ref{wdagger}) follows straightforwardly from (\ref{eq:48}),(\ref{eq:49}). One needs to use the identity 
$\delta_{s^{\pm},-s^{\prime\pm}+s^{\prime\prime\pm}}=\delta_{s^{\pm}+s^{\prime\pm},s^{\prime\prime\pm}}$ 
and the easily verifiable fact that $\alpha (s^{\pm}, s^{\prime\pm})$ is bilinear and antisymmetric in its
arguments.

\subsubsection{Unitary representation of finite gauge transformations.}
Since $Y^{\pm}$ are periodic scalar densities, under finite gauge transformations their holonomies 
transform in a similar manner to those of  the embedding momenta. Specifically, equation (\ref{eq:13*})
in conjunction with the periodicity of $\phi^{\pm}, Y^{\pm},s_{ext}^{\pm}$ and the various definitions of section 3.1,
imply that 
\begin{equation}\label{eq:46}
\alpha_{\phi^{\pm}}W_{s^{\pm}}[Y^{\pm}]\ :=\ W_{(\phi^{\pm})(s_{ext}^{\pm})\vert_{[0,2\pi]}}[Y^{\pm}].
\end{equation}
It is straightforward to see (either explicitly from equation (\ref{defalphass'}) or abstractly using the fact that the
periodicity of $\phi^{\pm}, Y^{\pm},s_{ext}^{\pm}$ 
implies that one is effectively restricting attention to diffeomorphisms,
graphs, charge networks and holonomies on $S^1$) that 
\begin{equation}
\alpha (s^{\pm},s^{\prime\pm})= \alpha (\phi^{\pm}(s^{\pm}_{ext})\vert_{[0,2\pi]},
\phi^{\pm}(s^{\prime\pm}_{ext})\vert_{[0,2\pi]}.
\label{alphadiff}
\end{equation}
Equations (\ref{eq:49}) and (\ref{alphadiff}) imply that the Hamiltonian flow of (\ref{eq:46}) induces 
an automorphism of the Weyl algebra of holonomies. 
Note also that equation (\ref{eq:47*}) is invariant under the action of this automorphism.
This directly implies that group of finite gauge transformations is  
unitarily represented in the quantum theory. Let these unitary 
operators be denoted, as in the embedding sector,
by ${\hat U}^{\pm}(\phi^{\pm})$. Their explicit action on the charge network basis can be defined from the GNS construction
to be
\begin{equation}\label{eq:54}
\begin{array}{lll}
\hat{U}^{\pm}(\phi^{\pm})W(s^{\pm})\ :=\ W((\phi^{\pm})(s^{\pm}_{ext})\vert_{[0,2\pi]}).\\
\end{array}
\end{equation}

\subsection{The kinematic Hilbert space.}
The kinematic Hilbert space ${\cal H}_{kin}$ is the product of the Hilbert spaces
${\cal H}^{\pm}_{kin}$ with 
\begin{equation}\label{hkinpm}
\mathcal{H}^{\pm}_{kin}\ =\ (\mathcal{H}_{E}^{\pm}\ \otimes\ \mathcal{H}_{M}^{\pm})
\end{equation}
so that 
\begin{equation}\label{eq:53*}
\mathcal{H}_{kin}\ =\ 
(\mathcal{H}_{E}^{+}\ \otimes\ \mathcal{H}_{M}^{+})\ \otimes\ (\mathcal{H}_{E}^{-}\ \otimes\ \mathcal{H}_{M}^{-}).
\end{equation}
$\mathcal{H}^{\pm}_{kin}$ is spanned by an orthonormal basis of equivalence classes of charge network states of the 
form $T_{s^{\pm}}\otimes W(s^{\prime\pm})$
with $s^{\pm}\ =\ \{\gamma(s^{\pm}),(k_{e_{1}^{\pm}}^{\pm},...,k_{e_{n^{\pm}}^{\pm}}^{\pm})\}$,
$s^{\prime\pm}\ =\ \{\gamma(s^{\prime\pm}),(l_{e_{1}^{\prime\pm}}^{\pm},...,l_{e_{m^{\pm}}^{\prime\pm}}^{\pm})\}$.

The results of the previous subsections show that ${\cal H}_{kin}$ supports
a *- representation of the *- algebras for the matter and embedding degrees of
freedom, as well as a unitary representation of finite gauge transformations.

Consider, as above, the state $T_{s^{\pm}}\otimes W(s^{\prime\pm})$.
The equivalence relation between charge networks is defined in
Definition 1, section 3.1. Using this equivalence, it is straightforward to see that we can always
choose $s^{\pm}, s^{\prime\pm}$ such that $\gamma(s^{\pm})=\gamma(s^{\prime\pm})$. Then each edge $e^{\pm}$ of
$\gamma (s^{\pm})$ is labelled by a pair of real charges $(k_e^{\pm},l_e^{\pm})$. Note that such a choice
graph and charge pairs is still not unique.
However it is easy to see that a unique choice can be made if we require that 
the  pairs of charges, $(k^{\pm}_{e^{\pm}}, l^{\pm}_{e^{\pm}})$, are such that no two consecutive edges are labelled
by the same pair of charges.
We shall denote this  unique labelling by ${\bf s}^{\pm}$ so that
\begin{equation}
{\bf s}^{\pm}:= \{\gamma {(\bf s}^{\pm}), (k^{\pm}_{e_1^{\pm}},l^{\pm}_{e_1^{\pm}}),...,(k^{\pm}_{e_{n^{\pm}}^{\pm}},l^{\pm}_{e_{n^{\pm}}^{\pm}})\},
\label{bfs}
\end{equation}
with
\begin{equation}
k_{e_{I^{\pm}}^{\pm}}\neq k_{e_{(I+1)^{\pm}}^{\pm}} \;{\rm or/and }\; l_{e_{I^{\pm}}^{\pm}}\neq l_{e_{(I+1)^{\pm}}^{\pm}}.
\label{unequalpair}
\end{equation}
The corresponding charge network state is denoted by $|{\bf s}^{\pm}\rangle$ so that
\begin{equation}
|{\bf s}^{\pm}\rangle =T_{s^{\pm}}\otimes W(s^{\prime\pm})
\label{s=tw}
\end{equation}
 with ${\bf s}^{\pm}$ defined from
$s^{\pm}, s^{\prime\pm}$ in the manner discussed above. It follows from (\ref{eq:56}) and (\ref{eq:54}) that
${\hat U}^{\pm}(\phi^{\pm})$ maps $|{\bf s}^{\pm}\rangle$ to a new charge network state. We denote the new (unique)
charge network label
by ${\bf s}^{\pm}_{\phi^{\pm}}$ so that
\begin{equation}
|{\bf s}^{\pm}_{\phi^{\pm}}\rangle := {\hat U}^{\pm}(\phi^{\pm})|{\bf s}^{\pm}\rangle .
\label{bfsphi}
\end{equation}

\section{Unitary representation of Dirac observables.}

\subsection{Exponentials of mode functions.}
Whereas $a_{(\pm )n}$ (\ref{eq:16}) depend on $Y^{\pm}(x)$, the basic operators of quantum theory are
the holonomies ${\hat W}(s^{\pm})$. As in LQG, the representation of the holonomy operators on ${\cal H}_{kin}$
is not regular enough to allow a definition of ${\hat Y}^{\pm}(x)$ via a ``shrinking of edges'' procedure \cite{alok}.
For example, let $s^{\pm}(t)$ be a 1 parameter family of charge networks such that $\gamma (s^{\pm}(t))$ has
non- vanishing unit charge on only one of its edges. Let this edge contain $x$ and let its  coordinate length
be $t$. Whereas, classically,
$Y^{\pm}(x)= \lim_{t\rightarrow 0}\frac{W(s^{\pm}(t))-1}{it}$, it is easy to check that, as in LQG, the corresponding
operators are not weakly continuous in $t$ and the limit cannot be defined on the charge network basis. This leads
to a regularization dependence in the definition of ${\hat a}_{(\pm )n}$ \cite{alok}. However, as we show below,
suitably defined exponential functions of $a_{(\pm )n},a^*_{(\pm )n}$ {\em can} be promoted to quantum operators
in a regularization/triangulation independent manner.
Let $q_n,p_n$ be the real and imaginary parts of $a_{(\pm )n}$ so that
\begin{equation}\label{eq:80}
\begin{array}{lll}
q_{(\pm)n} & = & \int_{S^{1}}Y^{\pm}(x) \cos(nX^{\pm}(x)),\\
\vspace*{0.1in}
p_{(\pm)n} & = & \int_{S^{1}}Y^{\pm}(x) \sin(nX^{\pm}(x)),\\
\end{array}
\end{equation}
and consider the functions
\begin{equation}\label{eq:81}
\begin{array}{lll}
e^{i\alpha q_{(\pm)n}}\ =\ e^{i\alpha\int_{S^{1}}Y^{\pm}(x) \cos(nX^{\pm}(x))}\\
\vspace*{0.1in}
e^{i\beta p_{(\pm)m}}\ =\ e^{i\beta\int_{S^{1}}Y^{\pm}(x) \sin(nX^{\pm}(x))}
\end{array}
\end{equation}
where $\alpha,\ \beta\ \in \mathbf{R}$. These functions can be promoted to quantum operators as follows.

Let $f(X^{\pm})$ be a smooth periodic {\em real }function
of $X^{\pm}$. Then $O^{\pm}_f: = \int_{S^1}Y^{\pm}(x)f(X^{\pm}(x))$ are functions on the phase space of PFT.
Next, restrict attention to the embedding sector Hilbert space ${\cal H}^{\pm}_E$ and consider the
operator valued (on  ${\cal H}^{\pm}_E$) function on the matter phase space,
$O^{\pm}_{\hat f}:= \int_{S^1}Y^{\pm}(x)f({\hat X}^{\pm}(x))$. Since charge network states are eigen states of
the embedding operator, we have that
\begin{equation}
O^{\pm}_{\hat f} T_{s^{\pm}}= (\sum_{i=1}^n f(\hbar k^{\pm}_{e_i^{\pm}})\int_{e_i^{\pm}} Y^{\pm}(x)) T_{s^{\pm}},
\label{ofhat}
\end{equation}
where $s^{\pm}= \{\gamma (s^{\pm}), (k^{\pm}_{e_1^{\pm}},..,k^{\pm}_{e_{n^{\pm}}^{\pm}}
)\}$ and that,
\begin{eqnarray}
e^{i O^{\pm}_{\hat f}} T_{s^{\pm}}
&=& e^{\sum_{i=1}^n f(\hbar k^{\pm}_{e_i^{\pm}})\int_{e_i^{\pm}} Y^{\pm}(x)}T_{s^{\pm}},\nonumber\\
&=& W(s^{\pm}_{f})[Y^{\pm}] T_{s^{\pm}},
\label{eofhat}
\end{eqnarray}
where $s^{\pm}_{f}:= \{\gamma (s^{\pm}), ( f(\hbar k^{\pm}_{e_1^{\pm}}),....,f(\hbar k^{\pm}_{e_{n^{\pm}}^{\pm}}))\}$.
Equation (\ref{eofhat}) implies that we can define the operators $\widehat {\exp {i {O}^{\pm}_{ f}}}$
corresponding to the
functions $\exp {i O^{\pm}_f}$ via their action  on the charge network states
$T_{s^{\pm}}\otimes W(s^{\prime\pm}) \in {\cal H}^{\pm}$:
\begin{equation}
{\widehat{(\exp {iO^{\pm}_{ f}}})}T_{s^{\pm}}\otimes W(s^{\prime\pm})
:={\hat W}(s^{\pm}_{f})T_{s^{\pm}}\otimes W(s^{\prime\pm}).
\label{hateof}
\end{equation}
Clearly, this is a manifestly regularization/triangulation independent definition.
Moreover, since $s^{\pm}_{f}$ is constructed from the embedding part of the charge network, and
since $f$ is periodic, it is straightforward to check that $\widehat{e^{i O^{\pm}_{ f}}}$ commute
with the unitary operators corresponding to finite gauge transformations. Hence $\widehat{e^{i O^{\pm}_{ f}}}$
are Dirac observables in quantum theory.
It is also easy to check that  
\begin{equation}
({\widehat{\exp i O^{\pm}_f}})^{\dagger}=({\widehat{\exp i O^{\pm}_f}})^{-1}=({\widehat{\exp i O^{\pm}_{-f}}})
\label{realitycondof}
\end{equation}
so that the classical reality 
conditions are implemented.

By setting 
$f$ to be the appropriate cosine (sine) function times $\alpha$ ($\beta$), we obtain
the operators corresponding to the functions in equation (\ref{eq:81}).
Clearly, these operators ($\forall \alpha, \beta \in \mathbf{R}, n>0$) form an over- complete
set of Dirac observables.

\subsection{Conformal Isometries.}
Regularization dependence also manifests in attempts to promote the generators of conformal isometries, 
$\Pi^{\pm}[U^{\pm}]$ (see equation
(\ref{eq:18}), to operators on ${\cal H}_{kin}$. Choosing exponentials of these observables only partially alleviates
this problem since (unlike the case of $a_{(\pm)n}$)  the resulting operator suffers from 
operator ordering problems stemming from the fact that $\{\Pi_{\pm}(x), U^{\pm}(X^{\pm}(x))\}\neq 0$.
Therefore, we focus on the Hamiltonian flows corresponding to finite conformal isometries.

The action of the Hamiltonian flows (corresponding to conformal isometries), $\alpha_{\phi_c^{\pm}}$, 
on $(X^{\pm}(x), Y^{\pm}(x))$ has been detailed in section 2.3.4. It remains to specify their action
on the embedding momenta, $\Pi_{\pm}(x)$. The information in this specification can equally well be seeded
in the action of $\alpha_{\phi_c^{\pm}}$ on the Hamiltonian flows $\alpha_{\phi^{\pm}}$ corresponding 
to finite gauge transformations  by virtue of the facts that (a) the constraints (\ref{eq:2pm})
are linear in the embedding momenta  and (b) this linear dependence is invertible by virtue of the non- degeneracy
condition {\bf (iv)} of section 2.3.1. Thus $\alpha_{\phi_c^{\pm}}$ are completely specified through
equations (\ref{phicalpha0}),(\ref{phicalpha1}),(\ref{phicalpha2}). Accordingly, we seek a unitary representation
of $\alpha_{\phi_c^{\pm}}$ by operators ${\hat V}(\phi_c^{\pm})$ such that 
${\hat V}^{\pm}(\phi_c^{\pm})$ act trivially on the matter sector, commute with the operators ${\hat U}^{+} (\phi^+)$ and
${\hat U}^{-} (\phi^-)$ which implement gauge transformations, and transform ${\hat X}^{\pm}(x)$ through
\begin{equation}
{\hat V}^{\pm}(\phi_c^{\pm}){\hat X}^{\pm}(x)({\hat V}^{\pm})^{\dagger}(\phi_c^{\pm}) = \phi_c^{\pm} ({\hat X}^{\pm}(x)),
\label{vhatx}
\end{equation}
while leaving ${\hat X}^{\mp}(x)$ invariant.

We define ${\hat V}^{\pm}(\phi_c^{\pm})$ to act trivially on the matter Hilbert spaces ${\cal H}^+_M,{\cal H}^-_M$ and on the $\mp$
embedding Hilbert space ${\cal H}^{\mp}_E$. The action of ${\hat V}^{\pm}(\phi_c^{\pm})$ on
${\cal H}^{\pm}_E$ is defined as follows. Let $s=\{ \gamma (s) (k^{\pm}_{e_1^{\pm}},...,k^{\pm}_{e_n^{\pm}})\}$
be a charge network. Define the charge networks $\phi_c^+(s^{+}), \phi_c^- (s^{-})$ by
\begin{equation}
\phi_c^{\pm}(s^{\pm}):= \{ \gamma (s^{\pm}), (\phi_c^{\pm}(k^{\pm}_{e_1^{\pm}}),...,\phi_c^{\pm}(k^{\pm}_{e_n^{\pm}}))\}.
\label{defphics}
\end{equation}
Then the action of ${\hat V}(\phi_c^{\pm})$ on the charge network state $T_{s^{\pm}}\in {\cal H}^{\pm}_E$ is defined to be
\begin{equation}
\hat{V}^{\pm}[\phi_{c}^{\pm}]T_{s^{\pm}}\ =\ T_{(\phi_{c}^{\pm})^{-1}(s^{\pm})}.
\label{vts}
\end{equation}
To reiterate, in the notation (\ref{defphics}) we have that \\ 
$(\phi_{c}^{\pm})^{-1}(s^{\pm})= \{\gamma (s^{\pm}), ((\phi_c^{\pm})^{-1}(k_{e_1^{\pm}}^{\pm}),...,(\phi_c^{\pm})^{-1}(k_{e_n^{\pm}}))\}$.

From equation (\ref{vts}), the invertibility of the functions $\phi_c^{\pm}$ (which follows from equation 
(\ref{phicnondeg})) and the inner product (\ref{eip}), it follows that \\
$\langle \hat{V}^{\pm}[\phi_{c}^{\pm}]T_{s^{\pm}}| \hat{V}^{\pm}[\phi_{c}^{\pm}]T_{s^{\prime\pm}}\rangle
=\langle T_{s^{\pm}}|T_{s^{\prime\pm}}\rangle $ $\forall s^{\pm}, s^{\prime\pm}$, thus showing unitarity.
It is also straightforward to check, using the quasiperiodicity of the functions $\phi_c^{\pm}$ (\ref{phicqp}), that
$\hat{V}^{\pm}[\phi_{c}^{\pm}]$ commutes with ${\hat U} (\phi^{\pm})$. By definition $\hat{V}^{\pm}[\phi_{c}^{\pm}]$ 
commutes with ${\hat U} (\phi^{\mp})$ and with the matter holonomies. Finally, it is easy to check that 
equation (\ref{vhatx}) holds when applied on any charge network state. Thus, our definition of 
$\hat{V}^{\pm}[\phi_{c}^{\pm}]$ provides a satisfactory definition of conformal isometries in quantum theory.

Note also that equation (\ref{vts}) implies that 
\begin{equation}
\hat{V}^{\pm}[\phi_{1c}^{\pm}]\hat{V}^{\pm}[\phi_{2c}^{\pm}]=\hat{V}^{\pm}[\phi_{2c}^{\pm}\circ\phi_{1c}^{\pm} ],
\label{phicrep}
\end{equation}
so that our definition of  $\hat{V}^{\pm}[\phi_{c}^{\pm}]$ implies 
an anomaly free representation (by right multiplication) of the group of conformal isometries.

\section{Physical state space by Group Averaging.}

Only gauge invariant states are physical so that physical states $\Psi$  must satisfy the condition 
${\hat U}^{\pm}(\phi^{\pm})\Psi = \Psi, \; \forall \phi^{\pm}$. A formal solution to this condition is to fix some
$|\psi\rangle \in {\cal H}_{kin}$ and set
$\Psi = \sum |\psi^{\prime}\rangle$ where the sum is over all distinct $|\psi^{\prime}\rangle$  which are
gauge related to $\psi$.  A mathematically precise implementation of this idea places the gauge invariant
states in the dual representation (corresponding to a formal sum over bras rather than kets) and goes by the name of
Group Averaging. The ``Group'' is that of gauge transformations and the ``Averaging'' corresponds to
the construction of a gauge invariant state from a kinematical one by giving meaning to the formal sum over
gauge related states. Specifically (for details see Reference \cite{ALM^2T}), the physical Hilbert space
can be constructed if there exists an anti-linear map $\eta$ from a dense subspace ${\cal D}$ of the kinematical
Hilbert space ${\cal H}_{kin}$,
to its algebraic dual ${\cal D}^*$, subject to certain requirements. The algebraic dual of  ${\cal D}$
is defined to be the space of linear mappings from ${\cal D}$ to the complex numbers. The requirements which $\eta$
needs to satisfy are as follows. Let $\psi_1, \psi_2 \in {\cal D}$, let ${\hat A}$ be a Dirac observable of interest
and let $\phi^{\pm}$ be a gauge transformation with ${\hat U}^{\pm}(\phi^{\pm})$ being its unitary implementation on ${\cal H}_{kin}$.
Let $\eta (\psi_1)\in {\cal D}^*$ denote the image of $\psi_1$ by $\eta$ and let $\eta (\psi_1)[\psi_2]$ denote the
complex number obtained by the action of $\eta (\psi_1)$ on $\psi_2$.
Then for all $\psi_1, \psi_2, {\hat A},\phi$ we require that \\
\noindent {\bf (1)} $\eta (\psi_1)[\psi_2]= \eta (\psi_1)[{\hat U} (\phi )\psi_2] $\\
\noindent {\bf (2)} $\eta (\psi_1)[\psi_2] =(\eta (\psi_2)[\psi_1])^*$, 
$\eta (\psi_1)[\psi_1] \geq 0$. \\
\noindent {\bf (3)} $\eta (\psi_1)[{\hat A}\psi_2]= \eta ({\hat A}^{\dagger}\psi_1)[\psi_2] $.\\

Here, we choose ${\cal D}$ to be the finite span of charge network states. Clearly due to the split of 
`$+$' and `$-$' structures, we may consider averaging maps ${\eta}^{\pm}$ on the dense sets 
${\cal D}^{\pm} \subset {\cal H}^{\pm}_{kin}$ separately. Here ${\cal D}^{\pm}$ is the finite span of states
of the form $|{\bf s}^{\pm}\rangle$ (see section 3.4 for the notation used here and below).
Define the action of $\eta^{\pm}$ on $|{\bf s}^{\pm}\rangle$ as
\begin{equation}\label{eq:120}
\begin{array}{lll}
\eta^{\pm} (|{\bf s}^{\pm}\rangle) & = & \eta_{[{\bf s}^{\pm}]}\sum_{{\bf s}^{\prime \pm}\in [{\bf s}^{\pm}]}
    <\ {\bf s}^{\prime\pm}|\\
\vspace*{0.1in}
& = & \eta_{[{\bf s}^{\pm}]}\sum_{\phi^{\pm}\in Diff_{[{\bf s}^{\pm}]}^{P}\mathbf{R}}<{\bf s}_{\phi^{\pm}}^{\pm}|,
\end{array}
\end{equation}
where
$[{\bf s}^{\pm}]\ =\ \{ {\bf s}^{\prime\pm}\vert {\bf s}^{\prime\pm}\ =\  {\bf s}_{\phi^{\pm}}^{\pm} \; {\rm for\ some\ }
\phi^{\pm}\}$, 
$Diff_{[\bf{s}^{\pm}]}^{P}\mathbf{R}$ is a set of gauge transformations such that
for each  ${\bf s}^{\prime\pm}\in\ [{\bf s}^{\pm}]$ there is precisely one gauge transformation  in the set which maps 
${\bf s}^{\pm}$ to ${\bf s}^{\prime\pm}$ and $\eta_{[{\bf s}^{\pm}]}$ is a positive real number depending only on the
gauge orbit $[{\bf s}^{\pm}]$.
The right hand side of equation (\ref{eq:120}) inherits an action on states in ${\cal D}$ from that of 
each of its summands. Due to the inner product (\ref{eip}), (\ref{eq:48}), only a finite number of terms in the sum
contribute so that 
$\eta^{\pm} (|{\bf s}^{\pm}\rangle)$ is indeed in ${\cal D}^*$. It is straightforward to see that 
$\eta^{\pm}$ satisfies the requirements {\bf (1)}, {\bf (2)} 
and that a positive definite 
 inner product $<,>_{phys}$ on the 
space $\eta^{\pm} {\cal D}^{\pm}$ can be defined through
\begin{equation}
<\eta^{\pm}(|{\bf s}_1^{\pm}\rangle), \eta^{\pm}(|{\bf s}_2^{\pm}\rangle)>_{phys}=
\eta^{\pm}(|{\bf s}_1^{\pm}\rangle) [|{\bf s}_2^{\pm}\rangle ]. 
\label{physip}
\end{equation}
If in addition, {\bf (3)} is also satisfied by $\eta^{\pm}$
the group averaging technique guarantees that 
the above inner product automatically implements  the adjointness conditions on the Dirac observables (which act by
dual action on ${\cal D}^{\pm *}$)
\footnote{Given $\Psi^{\pm} \in {\cal D}^{\pm *}$, $\psi^{\pm} \in {\cal D}^{\pm}$ and ${\hat A}_{\pm}$ such that 
${\hat A}_{\pm}^{\dagger}\psi^{\pm} \in {\cal D}^{\pm}$,
define ${\hat A}_{\pm}\Psi^{\pm}$ through 
${\hat A}_{\pm}\Psi^{\pm} [\psi^{\pm} ]:= \Psi^{\pm} [{\hat A}_{\pm}^{\dagger}\psi^{\pm}]$. This is the dual action.
\label{dualaction}
}
of section 4, 
by virtue of the fact that these conditions are  implemented on ${\cal H}_{kin}$.

In section 5.2 we use the requirement {\bf (3)} to constrain the positive real numbers 
$\eta_{[{\bf s}^{\pm}]}$ and thus bring down the enormous ambiguity in the inner product (\ref{physip}).
While the analysis can be done, in principle, for all of $\eta^{\pm} [{\cal D}^{\pm}]$, we shall, for simplicity, restrict
attention to a certain subspace of ${\cal D}^{\pm}$ which is left invariant by finite gauge transformations as well as 
the Dirac observables of section 4. In section 5.1 we define this `superselected' subspace.
Finally, in section 5.3 we display an irreducible representation of the operator algebra generated by the Dirac 
observables in conjunction with the gauge transformations.

\subsection{The chosen subspace of ${\cal D}$.}

Consider the charge network state $T_{s^{\pm}}\otimes W_{s^{\prime \pm}}$. Let $\gamma (s^{\pm})$  have $n^{\pm}$ edges
and the let the embedding charges on these edges be such that:\\
\noindent {\bf (a)} $\pm k^{\pm}_{e^{\pm}_{I^{\pm}}} \geq \pm k^{\pm}_{e^{\pm}_{(I^{\pm}-1)}} \; I^{\pm}=2,..,n^{\pm}$.\\
\noindent {\bf (b)} $\pm (k^{\pm}_{e^{\pm}_{n^{\pm}}} - k^{\pm}_{e^{\pm}_{1}} ) \leq \frac{2\pi}{\hbar}$.

These conditions are physically motivated. Condition {\bf (a)} is the  quantum analog of the classical non-degeneracy
condition {\bf (iv)} of section 2.3.1. Condition {\bf (b)} (together with {\bf (a)}) is the quantum version of the
classical property (implicit in the smoothness of $X^{\pm}(x)$ in conjunction with {\bf (ii), (iv)}) that the 
$X$ circle wraps around the $x$ circle once and only once. 

Henceforth we shall restrict attention to charge network states subject to {\bf (a)} and {\bf (b)}. Note that these
conditions define a superselection sector of ${\cal D}$ with respect to gauge transformations as well as the 
observables of section 4. We will refer to this subspace $\mathcal{D}_{({\bf a})({\bf b})}$.

\subsection{Commutativity of $\eta^{\pm}$ with Dirac observables.}

We focus on the `$+$' case and suppress the `$+$' superscripts wherever possible. 
The `$-$' case follows analogously.
We aim to restrict $\eta_{[{\bf s}]}$ by subjecting it to condition {\bf (3)} above.
We choose ${\hat A}:=\widehat{e^{iO^{+}_f}}$ (recall, from section 4.1, that  $O^{+}_f:= \int_{S^1}Y^+(x)f(X^+(x))$).
Thus we require that $\forall\ {\bf s}$,
\begin{equation}\label{eq:122}
\widehat{e^{i\int Y^{+}f(X^{+})}}\eta(|{\bf s}\rangle)\ =\ \eta(\widehat{e^{i\int Y^{+}f(X^{+})}}|{\bf s}\rangle)\ .
\end{equation}
As in equation (\ref{s=tw}) we set $|{\bf s}^{\pm}\rangle =T_{s^{\pm}}\otimes W(s^{\prime\pm})$. The equivalence
relation between charge network labels allows us, without loss of generality, to choose 
$\gamma ({\bf s})=\gamma ({s})=\gamma ({s^{\prime}})$. Equations (\ref{hateof}), (\ref{eq:49}), (\ref{defalphass'}) 
imply that
\begin{equation}
\widehat{e^{i\int Y^{+}f(X^{+})}}|{\bf s}\rangle =
{\hat W}_{s_f} |{\bf s}\rangle := e^{-\frac{i\hbar\alpha (s_f,s)}{2}}|{\bf s}(f)\rangle
\label{wsf}
\end{equation}
where
\begin{eqnarray}
{\bf s} &=& \{\gamma ({\bf s}), ((k_{e_1},l_{e_1}),..,(k_{e_n},l_{e_n}))\} \label{gammabfs}\\
{s} &=& \{\gamma ({\bf s}), (k_{e_1},..,k_{e_n})\} \\
{s_f} &=& \{\gamma ({\bf s}), (f(\hbar k_{e_1}),..,f(\hbar k_{e_n}))\}\\
{\bf s}(f) &=& \{\gamma ({\bf s}), ((k_{e_1},l_{e_1}+f(\hbar k_{e_1}),..,(k_{e_n},l_{e_n}+f(\hbar k_{e_n})\}\\
\label{gammabfsf}
\alpha(s_{f},s)& =& \sum_{I=1}^{n} f(\hbar k_{e_{I}})[l_{e_{I+1}}\ -\ l_{e_{I-1}}],\;e_0:= e_n,\ e_{n+1}:=e_1
\end{eqnarray}
Recall (see section 3.4) that ${\bf s}$ denotes the  unique labelling such that no two consecutive edges of 
$\gamma ({\bf s})$ have the same pair of charges. 
It is straightforward to see from equation (\ref{gammabfsf}) that for
$I=1,..,n-1$,
\begin{eqnarray}
k_{e_I}\neq k_{e_{I+1}} & \;{\rm or/and }& \;l_{e_I}\neq l_{e_{I+1}} \nonumber\\
\Rightarrow 
k_{e_I}\neq k_{e_{I+1}} &\;{\rm or/and }&  \;l_{e_I}+f(\hbar k_{e_I})\neq l_{e_{I+1}} +f(\hbar k_{e_{I+1}}). 
\label{unique}
\end{eqnarray}
Thus, consistent with the use of bold face notation (see section 3.4), 
${\bf s}(f)$ is also  the unique labelling such that 
no two consecutive edges of its underlying graph (also chosen to be
$\gamma ({\bf s})$) have the same pair of charges.

From footnote \ref{dualaction}
(\ref{wsf}), (\ref{alphadiff}), the fact that $\widehat{e^{i\int Y^{+}f(X^{+})}}$ commutes with gauge transformations,
 and(\ref{eq:120}), it follows that the right hand side of 
(\ref{eq:122}) is
\begin{equation}
\widehat{e^{i\int Y^{+}f(X^{+})}}\eta(|{\bf s}\rangle)
= \eta_{[{\bf s}]}e^{\frac{i\hbar\alpha (s_f,s)}{2}}
\sum_{\phi\in Diff_{[{\bf s}]}^{P}\mathbf{R}}<{\bf s(f)}_{\phi }|.
\end{equation}
and that the  left hand side of (\ref{eq:122}) is 
\begin{equation}
\eta(\widehat{e^{i\int Y^{+}f(X^{+})}}|{\bf s}\rangle)\
= \eta_{[{\bf s}(f)]}e^{\frac{i\hbar\alpha (s_f,s)}{2}}
\sum_{\phi\in Diff_{[{\bf s}(f)]}^{P}\mathbf{R}}<{\bf s}(f)_{\phi }|
\end{equation}
where $|{\bf s}(f)_{\phi }\rangle :={\hat U}(\phi)|{\bf s}(f)\rangle$.
Thus we need to impose 
\begin{equation}
\eta_{[{\bf s}]}\sum_{\phi\in Diff_{[{\bf s}]}^{P}\mathbf{R}}<{\bf s(f)}_{\phi }|
=\eta_{[{\bf s}(f)]}
\sum_{\phi\in Diff_{[{\bf s}(f)]}^{P}\mathbf{R}}<{\bf s}(f)_{\phi }|
\label{lhs=rhs}
\end{equation}
It is easy to see that we may choose 
\begin{equation}
Diff_{[{\bf s}]}^{P}\mathbf{R}=Diff_{[{\bf s}(f)]}^{P}\mathbf{R} .
\label{diffs=diffsf}
\end{equation}
This immediately follows from the fact that 
\begin{equation}
{\hat U}(\phi )\widehat{e^{iO^+_f}} |{\bf s}\rangle \neq \widehat{e^{iO^+_f}} |{\bf s}\rangle \;\;
{\rm iff} \;\; {\hat U}(\phi ) |{\bf s}\rangle \neq  |{\bf s}\rangle .\;\;
\label{invertcommute}
\end{equation}
Equation (\ref{invertcommute}) follows, in turn, from the invertibility of
$\widehat{e^{iO^+_f}}$ (\ref{realitycondof})  and its commutativity with ${\hat U}(\phi )$.
Equations (\ref{diffs=diffsf}), (\ref{lhs=rhs}) imply that 
\begin{equation}
\eta_{[{\bf s}]}=\eta_{[{\bf s}(f)]}.
\label{s=sf}
\end{equation}
Next, we analyse the consequences of the restriction (\ref{s=sf}). There are 2 cases:\\
\noindent Case 1: $[{\bf s}]$ is such that there exists some ${\bf s}\in [{\bf s}]$,
${\bf s} = \{\gamma ({\bf s}), ((k_{e_1}l_{e_1}),..,(k_{e_n}, l_{e_n}))\}$ with 
\begin{equation}
k_{e_1} < k_{e_2}<....<k_{e_n},\;\;(k_{e_n}- k_{e_1}) < 2\pi.
\label{case1}
\end{equation}
\noindent Case 2: The complement of Case 1.

We have analysed both cases. The analysis for Case 2 is quite involved and, in the interests of pedagogy,
we do not present it here. We shall focus only on Case 1 in this paper.
Accordingly, consider  ${\bf s}$ as in Case 1. We define 
${\tilde {\bf s}}$ to be the {\em embedding} charge network label which is 
obtained by dropping the matter charge labels from ${\bf s}$ so that $\gamma ({\tilde {\bf s}})=\gamma ({\bf s})$
with the edges of $\gamma ({\tilde {\bf s}})$ carrying the same embedding charges as in  ${\bf s}$.
Since ${\bf s}, {\bf s}(f)$  have the same embedding charges 
and the same underlying graph, we could equally well have obtained ${\tilde{\bf s}}$ by dropping the matter charge 
labels from ${\bf s}(f)$. 
Thus, using the ` ${\tilde{} }$ ' notation, we have that 
\begin{equation}
{\tilde {\bf s}} = {\tilde {\bf s}}(f) = (\gamma ({\bf s}), (k_{e_1},..,k_{e_n})).
\label{tildes}
\end{equation}

Next, note that we can always choose $f$ such that  
$f (\hbar k_{e_I})= -l_{e_I}, \; I=1,..,n$ so that ${\bf s}(f)$ has vanishing matter charges.
Clearly the property that all matter charges vanish is a gauge invariant statement.
This fact together with equation (\ref{tildes}) implies that 
 the set $[{\bf s}(f)]$ (with $f$ chosen as above) is isomorphic to the set of {\em embedding} charge networks
which are gauge equivalent to $\tilde{\bf s}$. Denoting the latter set by $[\tilde{\bf s}]$
we have, from equation (\ref{s=sf}) that 
$\eta_{[{\bf s}]}$ can only depend on the set $[\tilde{\bf s}]$. We denote this dependence 
through the notation
\begin{equation}
\eta_{[\tilde{\bf s}]} :=\eta_{[{\bf s}]}.
\label{etatilde}
\end{equation}

An identical analysis holds for the conformal isometry operators ${\hat V}(\phi_c)$. 
Equation (\ref{vts}) implies that
\begin{equation}
{\hat V}(\phi_c)|{\bf s}\rangle=: |\phi_c^{-1}({\bf s})\rangle.
\label{vbfs}
\end{equation}
${\bf s}$ is given by equations (\ref{gammabfs}), (\ref{case1}) and 
\begin{equation}
\phi_c^{-1}({\bf s})=\{ \gamma ({\bf s}), ((\phi_c^{-1}(k_{e_1}), l_{e_1}),..,(\phi_c^{-1}(k_{e_n}), l_{e_n}))\}.
\label{phicbfs}
\end{equation}
The invertibility of $\phi_c$ and its periodicity imply that $\phi_c^{-1}({\bf s})$ is the unique labelling 
such that no 2 consecutive edges have the same pairs of charges, and that the condition (\ref{case1}) is
preserved by the action of ${\hat V}(\phi_c)$.

Condition {\bf (3)} implies that, in obvious notation,
\begin{equation}
\eta_{[{\bf s}]}\sum_{\phi\in Diff_{[{\bf s}]}^{P}\mathbf{R}}<\phi_c^{-1}({\bf s})|{\hat U}^{\dagger}(\phi )
=\eta_{[\phi_c^{-1}({\bf s})]}
\sum_{\phi\in Diff_{[\phi_c^{-1}({\bf s})]}^{P}\mathbf{R}}<\phi_c^{-1}({\bf s})|{\hat U}^{\dagger}(\phi ).
\label{vlhs=vrhs}
\end{equation}
An argument identical to that in (\ref{invertcommute}) implies that 
$Diff_{[{\bf s}]}^{P}\mathbf{R}=Diff_{[\phi_c^{-1}({\bf s})]}^{P}\mathbf{R}$ so that 
\begin{equation}
\eta_{[{\bf s}]}= \eta_{[\phi_c^{-1}({\bf s})]}.
\label{etaphic}
\end{equation}
Clearly, given any pair of charge networks ${\bf s}_1, {\bf s}_2$ as in Case 1, with 
$\gamma ({\bf s}_1)=\gamma ({\bf s}_2)$
and with identical matter charges, there exists some $\phi_c$ such that 
$|{\bf s}_2\rangle= {\hat V}(\phi_c)|{\bf s}_1\rangle$. This, in conjunction with equations (\ref{etaphic}),
(\ref{etatilde}) implies
that $\eta_{[{\bf s}]}$ can only depend on the set of graphs $[\gamma ({\bf s})]$ which are obtained by the 
action of gauge transformations on $\gamma ({\bf s})$. Specifically, 
\begin{eqnarray}
[\gamma ({\bf s})] & = &\{\gamma^{\prime} \ 
{\rm s.t.} \ \exists  \phi \ {\rm s.t.}\ 
\gamma^{\prime}_{ext}= \phi (\gamma_{ext})\} 
\nonumber\\
\gamma & :=& \gamma ({\bf s}) ,
\end{eqnarray}
where we have used the notation defined in  section 3.1.
We denote this dependence of $\eta_{[{\bf s}]}$ through the notation
\begin{equation}
\eta_{[{\bf s}]}= \eta_{[\gamma ({\bf s})]}
\end{equation}
This completes our analysis of the rigging map.

\subsection{Cyclic representation}
We focus on the `$+$' sector of the algebra of operators and the `$+$' sector of the 
state space. As in section 5.2 we suppress `$+$' superscripts. The analysis for the `$-$'
case follows analogously. Cyclicity is defined with respect to an algebra of operators.
Here the putative generators of the algebra are the Dirac observables of section 4 and
the finite gauge transformations. As we shall see in section 6, neither does the commutator of two of the observables
of section 4.1 yield  a representation of the corresponding Poisson brackets nor  does their product yield
a representation of the appropriate Weyl algebra. As shown in section 6, 
the connection with classical theory is state dependent and 
only holds for semiclassical states 
(this is roughly similar to  what happens for area operators in LQG \cite{areaanomaly}).
Given this situation, we define the operator algebra in terms of the concrete
representation on ${\cal H}_{kin}$ ( or ${\cal H}_{phys}$) of the relevant operators rather than in terms
of abstract representations of classical structures.

Since the operators of section 4 as well as those for finite gauge transformations are unitary (and hence bounded),
the finite span of their products is well defined on ${\cal H}_{kin}$ so that it is possible to define the algebra
of operators generated by these elementary ones in terms of the action of elements of this algebra on ${\cal H}_{kin}$.
We denote this algebra of operators as ${\cal A}_{D,G}^{kin}$. In a similar manner, consider the algebra of operators
generated by the action of the Dirac observables of section 4 on ${\cal H}_{phys}$. Denote this algebra by 
${\cal A}_{D}$.

Fix a graph $\gamma$. Let ${\bf s}_{\gamma}$ be the set of charge networks such that
$\forall {\bf s}\in {\bf s}_{\gamma}, \gamma ({\bf s}) = \gamma$ and ${\bf s}$ satisfies
condition (\ref{case1}) on its embedding charges. Let $[{\bf s}_{\gamma}]$ be the set of 
charge networks which are gauge related to elements of ${\bf s}_{\gamma}$ i.e.
$\forall {\bf s}^{\prime} \in [{\bf s}_{\gamma}] \exists$ some gauge transformation 
$\phi$ and some ${\bf s} \in {\bf s}_{\gamma}$ such that ${\bf s}^{\prime}= {\bf s}_{\phi}$.
Finally, let ${\cal H}_{[\gamma]}$ be the (Cauchy completion of the)
finite span ${\cal D}_{[\gamma]}$ ($\subset\ \mathcal{D}_{(\bf{a})(\bf{b})}$)
of charge network states $| {\bf s}^{\prime}\rangle,\; {\bf s}^{\prime}\in [{\bf s}_{\gamma}]$.

The analysis of the preceding section shows that:\\
\noindent (1)
${\cal H}_{[\gamma]}\subset {\cal H}_{kin}$ provides a cyclic representation of the algebra ${\cal A}_{D,G}^{kin}$. 
Any  charge network state in ${\cal H}_{[\gamma]}$ is a cyclic state.\\
\noindent (2)
Group averaging of states in ${\cal D}_{[\gamma]}$ yields a cyclic  representation of
the algebra ${\cal A}_{D}^{phys}$ i.e. ${\cal A}_{D}^{phys}$ is represented cyclically  on
${\cal H}_{[\gamma],phys}\subset {\cal H}_{phys}$
where
${\cal H}_{[\gamma],phys}$ is the Cauchy completion (in the physical inner product) of $\eta( {\cal D}_{[\gamma]})$.
The group average of any charge network state in ${\cal D}_{[\gamma]}$ is a cyclic state.

Note that both ${\cal H}_{[\gamma]}$ and ${\cal H}_{[\gamma],phys}$ are {\em non- separable}.

\section{Semiclassical Issues.}
An exhaustive analysis of semiclassical states is outside the scope of this paper. Instead, we focus on two
issues related to semiclassicality. In section 6.1 we show that semiclassical states must be based on
suitably defined `weaves'. In section 6.2 we show that semiclassicality can be exhibited with respect to, at most,
a countable number of the mode function operators of section 4.1.

\subsection{Semiclassicality and Weaves.}
Recall that in LQG, states which exhibit semiclassical behaviour for spatial geometry operators are based on
graphs called weaves \cite{lqgweave}. Here the (flat) spacetime geometry is encoded in the behaviour of the
${\hat X}^{\pm}(x)$ operators. Hence we define the notion of a weave as follows. The embedding charge network
$s^{\pm}= \{\gamma (s^{\pm}), (k_{e_1^{\pm}}^{\pm},..,k_{e_{N^{\pm}}^{\pm}}^{\pm})\}$ will be called a {\em weave} iff the embedding charges satisfy
{\bf (a)},{\bf (b)} of section 5.1 together with $k_{e_N^{\pm}}^{\pm}- k_{e_1^{\pm}}^{\pm}\approx \pm2 \pi$ and iff $N>>1$.
This is, of course, not  a precise definition since  $k_{e_N^{\pm}}^{\pm}- k_{e_1^{\pm}}^{\pm}\approx 2 \pi$ and $ N>>1$.
are not precise statements. Nevertheless this `working' definition will suffice for our purposes.

Let $\psi^{\pm} \in {\cal H}_{kin}^{\pm}$ exhibit semiclassicality with respect to the $\pm$ sector observables
of section 4.1. Further, let $\psi^{\pm}$ be an eigen state of ${\hat X}^{\pm}(x)$ ( we shall relax this
assumption later) so that $\psi^{\pm}= T_{s^{\pm}}\otimes \psi_{M}^{\pm}$, $\psi_M^{\pm}\in {\cal H}_M^{\pm}$.
The analysis below is for the +-sector and can be trivially extended to the $-$-sector. In what follows we suppress the + superscript.
From equation (\ref{hateof}) it follows straightforwardly that
\begin{equation}
\langle [\widehat{e^{i\alpha q_m}},\widehat{e^{i\alpha p_m}}]\rangle
= -2i\sin (\frac{\alpha \beta \hbar}{2} f_{s,m})
\langle \widehat{e^{i\alpha q_m +i \beta p_m}}\rangle ,
\label{6.1.1}
\end{equation}
where
\begin{equation}
f_{s,m}:= \sum_{I=1}\cos(\hbar mk_{e_I})(\sin(\hbar mk_{e_{I+1}})-\sin(\hbar mk_{e_I})).
\label{fsm}
\end{equation}
where $k_{e_{N+1}}:= k_{e_1}$. In order to write (\ref{fsm}) in a more useful form, we define the following:
\begin{eqnarray}\label{deltadef}
\Delta k_{e_I}&:= & k_{e_{I+1}}- k_{e_I},\;\; I = 1,.., N-1 \\
\Delta k_{e_N}&:= & k_{e_1}- k_{e_N} + \frac{2\pi}{\hbar} .
\end{eqnarray}
Rearranging terms
in (\ref{fsm})  and using  standard trigonometric identities we obtain that
\begin{equation}
f_{s,m}
=  \sum_{I=1}^N \sin (\hbar m \Delta  k_{e_I}).
\label{sindeltak}
\end{equation}

Since $\psi$ is semiclassical we assume that, for some classical data $(q_m, p_m)$,
\begin{equation}
\langle \widehat{e^{i\alpha q_m +i \beta p_m}}\rangle \approx  {e^{i\alpha q_m +i \beta p_m}},
\label{6.1.2}
\end{equation}
and we require
that  as $\hbar \rightarrow 0$
\begin{equation}
\langle [\widehat{e^{i\alpha q_m}},\widehat{e^{i\alpha p_m}}]\rangle
\rightarrow i\hbar \{ e^{i\alpha q_m},e^{i\beta p_m}\}
\label{6.1.3}
\end{equation}
where the Poisson bracket evaluates to
\begin{equation}
\{ e^{i\alpha q_m},e^{i\beta p_m}\}= -\alpha \beta 2\pi m e^{i\alpha q_m +i\beta p_m}.
\label{6.1.4}
\end{equation}
Equations (\ref{6.1.1})- (\ref{6.1.4}) imply that to leading order in $\hbar$
\begin{equation}
f_{s,m=1} \approx 2\pi  .
\label{6.1.5}
\end{equation}
Note that the eigen values of the embedding operators are in terms of  
\begin{equation}
k_I:= \hbar k_{e_I}
\label{defki}
\end{equation}
so that in the $\hbar \rightarrow 0$ (classical) limit, $k_I$ does not vanish (except when $k_{e_I}=0$).
Hence, we investigate the conditions imposed on $s$ by the requirement
\begin{equation}
|2\pi  - \sum_{I=1}^N \sin (\Delta k_I) | < \epsilon , \;\;\; \epsilon <<1 .
\label{6.1.6}
\end{equation}
where, similar to (\ref{deltadef}) we have defined
\begin{eqnarray}\label{defdelta}
\Delta k_{I}&:= & k_{{I+1}}- k_{I},\;\; I = 1,.., N-1 \\
\Delta k_{N}&:= & k_{1}- k_{N} + 2\pi.
\end{eqnarray}
Note that conditions ${\bf (a)},{\bf (b)}$ of section 5.1 imply that 
\begin{equation}
\Delta k_I \geq  0, \;\;\; \sum_{I=1}^N\Delta k_I = 2\pi .
\label{sumdeltak}
\end{equation}

Intuitively, since
$|\frac{\sin x}{x}|\leq 1$ and $=1$ at $x=0$, equations (\ref{6.1.6}), (\ref{sumdeltak}) lead us to expect
that $\Delta k_I, I=1,..,N$ should be small. That this is indeed 
the case is shown in Lemmas 1- 3  in the Appendix. Clearly, 
the fact that $ \Delta k_I \rightarrow 0$ as $\epsilon \rightarrow 0$ (see Appendix)

implies that $s$ is a weave.
Thus, we have shown that any kinematic semiclassical state which is an eigen state of the embedding operators must be
based on a weave.

Next, consider an arbitrary kinematic state
$|\psi \rangle = \sum a_i |s_i>\otimes |\psi_{iM}\rangle$
where $a_i$ are complex coefficients, $|s_i\rangle$ are an orthonormal set of embedding charge network states
 and $|\psi_{iM}\rangle \in {\cal H}_M$. In order that this state satisfies equation (\ref{6.1.3}), it 
turns out that $|\psi \rangle$ must be peaked around $s_i$ such that $s_i$ are weaves. 
This is shown in Lemma 4
of the Appendix. 

Finally, consider an arbitrary physical state. Such a state is a linear combination of averages over
embedding eigen states. Lemma 5 shows that such a state is peaked around averages of embedding eigen states
which are based on weaves.

\subsection{ Semiclassicality and mode function operators: a no- go result.}
We show that no states exist which are semiclassical with respect to the uncountable set of 
operators $\{\widehat{e^{i\alpha q_m}},\widehat{e^{i\beta p_m}}, |\alpha -\alpha_0|<\epsilon,
|\beta - \beta_0|<\delta\}$ for any fixed $m, \alpha_0, \beta_0$ and any $\epsilon, \delta >0$.
First, consider states $|\psi\rangle$ which are embedding eigen states so that 
$|\psi\rangle = |s\rangle\otimes |\psi_M\rangle$. Here $s$ is an embedding charge network and 
$|\psi_M\rangle\in {\cal H}_M$ can expanded as $|\psi_M\rangle= \sum_{r}b_r |s^{\prime}_r\rangle$
where $\{|s^{\prime}_r\rangle\}$ is a countable set of orthonormal matter charge networks.

The operators $\widehat{e^{i\alpha q_m}},\widehat{e^{i\beta p_m}}$ act by changing the matter charge labels
by sines and cosines of ($m$ times) the  embedding charges (see (\ref{hateof}). Consider the set 
$L$ of all matter charges on $s_r \forall r$ and construct the set $\Delta L$ of differences between 
all pairs of elements of $L$ i.e. $\Delta L :=\{l-l^{\prime}\forall l,l^{\prime} \in L\}$. Let 
$k_e, e \subset \gamma (s)$ be such that $\cos m\hbar k_e \neq 0$. Then, in any neighbourhood of
$\alpha_0$ we can choose uncountably many $\alpha$ such that $\alpha \cos m\hbar k_e \notin \Delta L$.
Clearly for such $\alpha$ we have that $\langle\widehat{e^{i\alpha q_m}}\rangle=0$. If 
$\cos m\hbar k_e =0$ we can repeat the same argument with $\sin m\hbar k_e$ and conclude that 
$\langle\widehat{e^{i\beta p_m}}\rangle=0$ for uncountable many $\beta$ near $\beta_0$. Clearly, such
behaviour is far from semiclassical. This argument can be suitably generalised for arbitrary states
in ${\cal H}_{kin}$ as well as in ${\cal H}_{phys}$. The relevant material is in Lemma 6 and Lemma 7 of
Appendix B.

\section{Two open issues and their resolution.}
Before we conclude this paper, a couple of points remain which we have not addressed as yet.
First, it still remains to enforce {\bf(ii)}, section 2.3.1 in order to ensure that the spatial topology is 
a circle. Second, we need to take care of the zero modes by imposing equation (\ref{p=0}) in quantum 
theory and show that the results of section 6 continue to hold after this is done. We address these points
in sections 7.1 and 7.2 below.

\subsection{Identifying $2\pi$ shifted embeddings}
Although the spatial inertial co-ordinate $X$ ranges over $(-\infty, \infty)$, we need to identify
$X\sim\ X\ +\ 2\pi$ in accordance with the discussion in section 2.3.1. 
Condition {\bf (ii)}, section 2.3.1 states that 
two embeddings $(X_{1},T_{1})$, $(X_{2}, T_{2})$ are equivalent if the following conditions are satisfied:

\begin{equation}\label{eq:149}
\begin{array}{lll}
X_{1}^{+}(x) = X_{2}^{+}(x) + 2m\pi\ \forall\ x\in\ [0,2\pi],\\
X_{1}^{-}(x) = X_{2}^{-}(x) - 2m\pi\ \forall\ x\in\ [0,2\pi].\\
\end{array}
\end{equation}
We now show that this equivalence has already been taken care of at the physical state-space level.
Let
\begin{equation}\label{eq:149.1}
\begin{array}{lll}
{\bf s}^{+}\ =\ \{\ \gamma({\bf s}^{+}), (k_{e_{1}^{+}}^{+},...,k_{e_{N}^{+}}^{+}), (l_{e_{1}^{+}}^{+},...,l_{e_{N}^{+}}^{+})\ \}\\
\vspace*{0.1in}
{\bf s}^{-}\ =\ \{\ \gamma({\bf s}^{-}), (k_{e_{1}^{-}}^{-},...,k_{e_{M}^{-}}^{-}), (l_{e_{1}^{-}}^{-},...,l_{e_{M}^{-}}^{-})\}
\end{array}
\end{equation}
The identification (\ref{eq:149.1}) in the classical theory implies the following equivalence condition in 
quantum theory:
\begin{equation}\label{eq:150}
\vert{\bf s}^{+}\rangle\otimes \vert{\bf s}^{-}\rangle\sim\ \vert{\bf s}^{+}_{2\pi m}\rangle\otimes 
\vert{\bf s}^{-}_{2\pi m}\rangle
\end{equation}
where,
\begin{equation}\label{eq:151}
\begin{array}{lll}
{\bf s}^{+}_{2\pi m}\ =\ \{\ \gamma({\bf s}^{+}), (k_{e_{1}^{+}}^{+}+2m\pi,...,k_{e_{N}^{+}}^{+}+2m\pi), (l_{e_{1}^{+}}^{+},...,l_{e_{N}^{+}}^{+})\ \},\\
{\bf s}^{-}_{-2\pi m}\ =\ \{\ \gamma({\bf s}^{-}), (k_{e_{1}^{-}}^{-}-2m\pi,...,k_{e_{M}^{-}}^{-}-2m\pi), (l_{e_{1}^{-}}^{-},...,l_{e_{M}^{-}}^{-})\ \}.
\end{array}
\end{equation}

Next, note that for any integer $m$, there exist gauge transformations $\phi^{\pm}_{(m)}$ 
such that 
$\phi^{\pm}_{(m)}\cdot {\bf s}^{\pm}= 
\{\gamma({\bf s}^{\pm}),\ (k_{e_{1}^{\pm}}^{\pm} \pm 2m\pi,...,k_{e_{N}^{\pm}}^{\pm} \pm 2m\pi),
\ (l_{e_{1}^{\pm}}^{\pm},...,l_{e_{N}^{\pm}}^{\pm})\}$.
Thus $\vert{\bf s}^{\pm}\rangle$ and $\vert{\bf s}_{\pm 2\pi m}^{\pm}\rangle$ are gauge related so that 
\begin{equation}\label{eq:153}
\eta^{\pm}(\vert{\bf s}^{\pm}\rangle)\ =\ \eta^{\pm}(\vert{\bf s}_{\pm 2\pi m}^{\pm}\rangle),
\end{equation}
\begin{equation}\label{eq:154}
\Rightarrow
\eta^{+}(\vert{\bf s}^{+}\rangle)\otimes\ \eta^{-}(\vert{\bf s}^{-}\rangle)\ =\ 
\eta^{+}(\vert{\bf s}_{2\pi m}^{+}\rangle)\otimes\ \eta^{-}(\vert{\bf s}_{-2\pi m}^{-}\rangle).
\end{equation}
Equation (\ref{eq:154}) shows that 
 the identification of $2\pi$-shifted embeddings is \emph{subsumed} by the 
identification of embeddings related by gauge transformations.

\subsection{Taking care of the zero mode in quantum theory.}
In section 7.2.1 we impose the condition $p=0$ (see equation (\ref{p=0})) by appropriate group averaging. In
section 7.2.2 we show that this does not alter the conclusions of section 6.

\subsubsection{Imposition of $p=0$ by averaging.}

The conditions $\int_{S^{1}}Y^{\pm}=0$ of equation (\ref{p=0}) are equivalent to the conditions
$e^{i\lambda^{\pm} \int_{S^{1}}Y^{\pm}}= 1,\ \forall \lambda^{\pm}$. The latter can be imposed by group 
averaging with respect to the operators 
$\widehat{e^{i\lambda^{\pm} \int_{S^{1}}Y^{\pm}}}$. Let $s^{\pm}_{\lambda^{\pm}}$ be matter charge networks with a 
single edge $e^{\pm}= [0,2\pi ]$ labelled by the charge $\lambda^{\pm}$ i.e.
$s^{\pm}_{\lambda^{\pm}}= \{\gamma (s^{\pm}_{\lambda^{\pm}})= [0,2\pi ], l^{\pm}_{e^{\pm}}= \lambda^{\pm}\}$. Clearly,
we have that $\widehat{e^{i\lambda^{\pm} \int_{S^{1}}Y^{\pm}}}= \hat{W}(s_{\lambda^{\pm}}^{\pm})$.

It is easy to see that $U^{\pm}(\phi^{\pm}) \hat{W}(s_{\lambda^{\pm}}^{\pm})U^{\pm}((\phi^{\pm})^{-1})\ =\ \hat{W}(s_{\lambda^{\pm}}^{\pm})$. Thus we can average over the transformations generated by the zero-mode constraint before or after averaging over the group of gauge transformations. Since we have already averaged over the Virasoro group, we solve the zero-mode constraint by
defining a Rigging map 
$\overline{\eta}^{\pm} : \eta^{\pm}(\mathcal{D}_{(a)(b)}^{\pm})\rightarrow \eta^{\pm}(\mathcal{D}_{(a)(b)}^{\pm})^{*}$. 
Recall that $\mathcal{D}_{(a)(b)}^{\pm}$ (defined in section 5.1) is the finite span of charge networks subject to 
the conditions $({\bf a}), ({\bf b})$ of section 5.1. 
$\eta^{\pm}(\mathcal{D}_{(a)(b)}^{\pm})^{*}$ is the algebraic dual of $\eta^{\pm}(\mathcal{D}_{(a)(b)}^{\pm})$. 
Before defining $\overline{\eta}^{\pm}$, note that,
\begin{equation}\label{eq:155}
\hat{W}(s_{\lambda^{\pm}}^{\pm})\vert {\bf s}^{\pm}\rangle\ =:\ \vert {\bf s}^{\pm}_{\lambda^{\pm}}\rangle
\end{equation}
where ${\bf s}^{\pm}_{\lambda^{\pm}}$ is obtained from ${\bf s}^{\pm}=\{ \gamma({\bf s})^{\pm},\vec{k}^{\pm},\vec{l}^{\pm}\}$ by adding $\lambda^{\pm}$ to all the matter charges.
We now define,
\begin{equation}\label{eq:156}
\begin{array}{lll}
\overline{\eta}^{\pm}(\eta^{\pm}(\vert {\bf s}^{\pm}\rangle)) = 
\overline{\eta}_{[[{\bf s}^{\pm}]]_{0}}\eta_{[{\bf s}^{\pm}]}
(\bigoplus _{\lambda^{\pm}\in \mathbf{R}}\sum_{\phi^{\pm}\in Diff_{[\gamma({\bf s}^{\pm})]}^{P}\mathbf{R}} 
\langle ({\bf s}^{\pm}_{\phi^{\pm}})_{\lambda^{\pm}} |
\end{array}
\end{equation}
The equivalence class $[[{\bf s}^{\pm}]]_{0}$ is defined via following relation.\\
$[{\bf s}^{\pm}]\sim [{\bf s}_{1}^{\pm}]$ iff for any $\{\gamma({\bf s}^{\pm}),\vec{k}^{\pm},\vec{l}^{\pm}\} \in [{\bf s}^{\pm}]$, $\exists$ $(\{\gamma({\bf s}^{\pm}),\vec{k}^{\pm},\vec{l}^{\pm}+\lambda_{\pm}\} \in [{\bf s}^{\pm}_{1}]$ for some $\lambda_{\pm}\in \mathbf{R}$.\\
\hspace*{0.2in} Once again the ambiguity in the rigging map contained in $\overline{\eta}_{[[{\bf s}^{\pm}]]_{0}}$ can be reduced by demanding that $\overline{\eta}^{\pm}$ commutes with the observables. It can be checked that for the super-selected sector of $\mathcal{H}_{phy}$ defined in section 5.2, we have $\overline{\eta}_{[[{\bf s}^{\pm}]]_{0}}\ =\ \overline{\eta}_{[\gamma({\bf s}^{\pm})]}$. 
Setting ${\tilde {\eta}}_{[\gamma({\bf s}^{\pm})]}:=
\overline{\eta}_{[\gamma({\bf s}^{\pm})]}\eta_{[\gamma({\bf s}^{\pm})]}$, we have that the 
inner product on $\overline{\eta}^{\pm}(\mathcal{D}^{\pm\ ss}_{phy})$ is given by,
\begin{equation}\label{eq:157}
\langle\overline{\eta}^{\pm}(\eta^{\pm}(\vert{\bf s}^{\pm}\rangle)\vert \overline{\eta}^{\pm}(\eta^{\pm}(\vert{\bf s}_{1}^{\pm}\rangle)\rangle\ =\ {\tilde {\eta}}_{[\gamma({\bf s}^{\pm})]}
\ \bigoplus_{\lambda^{\pm}} \Big(\ \eta^{\pm}(\vert{\bf s}^{\pm}\rangle)[\vert {\bf s}_{1,\lambda^{\pm}}\rangle]\Big),
\end{equation}

\subsubsection{Semiclassical Issues.}

Since the zero mode operator $\hat{W}(s_{\lambda^{\pm}}^{\pm})$ leaves the embedding part of the states in 
${\cal H}_{kin}$ and ${\cal H}_{phys}$ untouched, it is easy to see that the proofs of section 6.1 and appendix A
still apply after the zero mode averaging is done. Thus, semiclassical states which satisfy the $p=0$ constraint 
are necessarily based on weaves.

It is also straightforward to see that the results of section 6.2  apply after zero mode
group averaging. While the line of argument is roughly similar to that in section 6.2 and appendix B, there
are some differences.
In the interests of brevity, we provide only a skeleton of the argument below. As usual we shall suppress the $\pm$
superscripts.

The averaging with respect to $\overline{\eta}$ slightly complicates matters because there is an additional sum
over matter charge networks wherein  matter charges associated with  charge network states are all incremented
by the same amount. As a result, it is necessary to consider pairs of edges subject to conditions on their
embedding charges. This is in contrast to the role of single edges (with cosines or sines of ($\hbar$ times) their
embedding charges being non- vanishing) in the arguments of section 6.1 and appendix B. Specifically, 
consider a state decomposition defined in terms of embedding charge networks $s_j$ as in 
equations (\ref{lincomb}) and (\ref{l3.1}). Separate the values taken by the index $j$ into a set $C_1$ and
its complement, $C_2$, where $j\in C_1$ iff for fixed $m$, 
there exist a pair of edges $e_I(j), e_J(j) \in \gamma (s_j)$ such that 
$\cos m\hbar k_{e_I(j)}\neq \cos m\hbar k_{e_J(j)}$. 

Next, with a slight abuse of notation,
 for each $j\in C_1$ fix
a pair of edges  $e_I(j), e_J(j) \in \gamma (s_j)$ such that 
$\cos m\hbar k_{e_I(j)}\neq \cos m\hbar k_{e_J(j)}$. As in appendix B, define $\Delta L$ to be the set of differences
of all matter charges which occur in the expansions (\ref{lincomb}), (\ref{l3.1}), (\ref{l6.1}).
Also define $\Delta^2L$ to be the set of all differences between pairs of elements of $\Delta L$.
For each $j\in C_1$ define $\Delta^2L_j$ to be the set of elements obtained by dividing each element of 
$\Delta^2L$ by $\cos m\hbar k_{e_I(j)}-\cos m\hbar k_{e_J(j)}$. Let 
$\Delta^2L_{C_1}:= \cup_{j\in C_1}\Delta^2L_j$.
The set $\Delta^2L_{C_1}$ is countable so that there are uncountably many $\alpha$ in any neighbourhood of $\alpha_0$
such that $\alpha \notin \Delta^2L_{C_1}$. It can then be checked that 
$\langle \widehat{e^{i\alpha q_m}}\rangle$ obtains contributions only from terms labelled by $j\in C_2$.

Finally, we show that such terms are of negligible measure.
Note that for $j\in C_2$ we have that  $\cos m\hbar k_{e_I(j)}= \cos m\hbar k_{e_J(j)}$ for any pair of edges
$e_I(j), e_J(j) \in \gamma (s_j)$. It is then straightforward to see that  for such $j$, the function
$f_{s_j,m}$ (defined by equations (\ref{fsim}), (\ref{fsm})) vanishes identically. Then the arguments of section 6.1
and appendix A imply that the contribution from $j\in C_2$ must be negligible for semiclassicality to hold.

Similar arguments can be made for 
$\langle \widehat{e^{i\beta p_m}}\rangle$ by replacing cosines with sines in the above argument.

\section{Discussion of results and open issues.}
In this work, we constructed a quantization of PFT similar to that used in LQG. Quantum states are in correspondence
with graphs (i.e. collections of edges) in the spatial manifold. The edges of these graphs are labelled by 
a set of real valued embedding and matter charges. These charge network states are analogs of the spin network states
in LQG. There, however, the labels are integer valued. Such a labelling is also, in principle, possible here.
Had the holonomies of section 3 been based on charge networks with embedding charges which were integer
multiples of $\frac{2\pi}{L}$ for some fixed integer $L$ and  matter charges which were also integer multiples
of some appropriate dimensionful unit, such holonomies would still separate points in phase space by virtue of the fact 
that
they were based on arbitrary graphs (this is similar to what happens in LQG). Such a choice would lead to states with 
integer valued charges. However it is not clear if (a large enough subset of) the Dirac observables of section 4
preserve the space spanned by these integer-charge network states. It would be useful to investigate this issue in detail.

The polymer quantization of the embedding variables replaces the classical (flat) spacetime continuum with 
a discrete structure consisting of a countable set of points. This can be seen as follows.
The canonical data $X^{\pm}(x)$ is a map from $S^1$ into the flat spacetime $(S^1\times R, \eta )$ and embeds the former
into the latter as a spatial Cauchy slice. Any gauge transformation generated by the constraints maps this 
data to new embedding data which, in turn, define a new Cauchy slice in the flat spacetime. In particular, the action of
the one parameter family of gauge transformations generated by smearing the constraints with some choice
of ``lapse-shift'' type functions $N^A$ (see section 2) generates a foliation of $(S^1\times R, \eta )$. Consider 
the image set in $(S^1\times R, \eta )$ of the set of all embeddings which are gauge related to a given one.
From the above discussion it follows that this image set is exactly the flat spacetime $( S^1\times R, \eta )$ itself.
Next, consider the corresponding quantum structures. Any charge network state is an eigen state of ${\hat X}^{\pm}(x)$. 
Consider a charge network state,
$|{\bf s}^+\rangle\otimes |{\bf s}^-\rangle$ with $|{\bf s}^{\pm}\rangle= T_{s^{\pm}}\otimes W_{s{\prime\pm}}$,
where $s^{\pm}$ satisfy the conditions ${\bf (a)},{\bf (b)}$ of section 5.1.
From equation (\ref{eq:33})- (\ref{eq:33b}) it follows that 
the set of eigen values 
$\lambda_{x,s^{\pm}}$ for all $x\in [0,2\pi]$ describes a finite set of points on a spacelike Cauchy surface in 
$(S^1\times R, \eta )$. These points have light cone coordinates $(X^+, X^-) = (\lambda_{x,s^{+}}, \lambda_{x,s^{-}})$.
The action of any gauge transformation on such a charge network state yields another charge network state
whose eigen values lie, once again, on a Cauchy slice in $(S^1\times R, \eta )$. From equation (\ref{eq:56})
it follows that the set of  eigen values for all possible gauge related charge network states is countable
and defines a corresponding set of points in $(S^1\times R, \eta )$. The gauge invariant state obtained by
group averaging a charge network state is a sum over all distinct gauge equivalent states and hence contains
the elements of this discrete structure. The discrete structure is a good approximant  of the continuum
spacetime
 $(S^1\times R, \eta )$ for charge networks with a large number of embedding charges i.e. for weave states.
Thus, it is not surprising that semiclassicality requires states to be based on weaves as in section 6.1 and appendix
A. 

In contrast to the embedding charges, the matter charges do not have a direct physical interpretation because
charge network states are not eigen states
of the 
matter holonomies.
As a tentative, provisional
interpretation we choose to think of them, rather imprecisely, as measuring excitations of the matter.
Since, on the constraint surface, the classical data $(X^{\pm}(x), Y^{\pm}(x))$ correspond to free scalar field 
data $Y^{\pm}(x)$ on the slice $(X^+(x), X^-(x))$ in flat spacetime, we interpret  a charge network state 
$|{\bf s}^+\rangle\otimes |{\bf s}^-\rangle \in{\cal H}_{kin}$ as 
specifying excitations of matter on the discretized ``quantum'' slice specified by the embedding charges.
The action of a gauge transformation on a charge network state can then be interpreted as evolving the 
matter excitations on the `initial' quantum slice specified by this state to the new one specified by the 
gauge related charge network state. Since the physical state obtained as the 
group average of a charge network state contains all distinct
gauge related states, it follows that such a physical state may be interpreted, roughly, as a ``history''.
It may be useful to attempt an interpretation of  physical states in LQG along these lines.

An over- complete set of Dirac observables corresponding to exponential functions of the standard annihilation- creation
modes of free scalar field theory are represented as (unitary) operators in the polymer representation.
Note that in contrast to the assumption of Reference \cite{ALM^2T}, here the commutator between two such operators
does not close as in the case of Weyl algebras. Indeed, as shown in section 6.1, the commutator only approximates the
corresponding Poisson bracket for semiclassical states based on weaves. This underlines the fact that in 
a general covariant theory involving spacetime geometry, classical structures are typically not approximated
in the $\hbar \rightarrow 0$ limit unless it is possible to coarse grain/smoothen away the underlying discreteness
of the quantum spacetime. Nevertheless the action of the basic Dirac observables is well defined and there is no
obstruction to the quantization procedure.

The results of section 6.2 imply that semiclassical analysis requires a choice of a countable subset of
these observables. One possibility is to choose, for each $n$,  a pair $\alpha, \beta << \frac{1}{\sqrt \hbar}$
and define the approximants to ${\hat q}_n, {\hat p}_n$ by 
$\frac{\widehat{e^{i\alpha q_n}}-\widehat{e^{-i\alpha q_n}}}{2i\alpha}$,
$\frac{\widehat{e^{i\beta p_n}}-\widehat{e^{-i\beta p_n}}}{2i\beta}$. 
However, there is no natural choice
of $\alpha, \beta$ and so, while the quantization constructed in this paper is free of the ``triangularization''
choices which occur in the definition of the quantum dynamics of LQG, an element of choice does appear 
when semiclassical issues are confronted. Note, however, that the results of section 6.1 indicate that 
any physical semiclassical state necessarily has an associated (gauge invariant) structure, namely that of a weave.
\footnote{Note that in contract to the weaves of Reference \cite{lqgweave} which approximate a spatial geometry, 
here it is the (flat) {\em spacetime} geometry which is being approximated by virtue of the discussion in the 
second paragraph of this section.}
The ``spacing'' of the weave (i.e. $\hbar \Delta k_I$ of section 6.1 and the Appendix A) provides a natural scale for
$\alpha, \beta$. Thus, our viewpoint is that since choices of Dirac observables can be tied (however tenuously) to 
structures already present in the semiclassical states, ambiguities (if present) in definitions of the 
quantum dynamics are more worrying because quantum dynamics is defined for all states, not only semiclassical ones.

The above discussion naturally brings us to the efficacy of polymer PFT as a toy model for LQG. We believe that 
the quantization provided here is a useful testing ground for proposed definitions of quantum dynamics
in canonical LQG. It would be of interest to construct the quantum dynamics of the model along the 
lines of Reference \cite{ttalqg} and compare the resulting physical Hilbert space with the one
considered here. Proposals for examining semiclassical issues \cite{ttcoherent,aashadow} may also be 
tested here. 
One of the outstanding problems in LQG \cite{megraviton,carlopropagator} is the relation between
states in LQG and the Fock states of perturbative gravity. Since PFT admits a Fock quantization \cite{karelqm,charlie}
equivalent to the standard flat spacetime free scalar field Fock representation, one may enquire as to how Fock states
arise from the polymer Hilbert space.
Since the results of section 6.2 suggest that  the  operators corresponding to exponentials of mode functions
do not possess the requisite continuity for the annihilation- creation
modes themselves to be defined as operators, it is difficult to identify Fock states in terms of their
properties with respect to the action of the annihilation- creation operators. However, as a first step,
it may be  possible to identify candidate states corresponding to the Fock vacuum by using the Poincare invariance of 
the latter. Specifically, since the operators corresponding to finite Poincare transformations are available (as a
subset of the conformal isometry operators of section 4), one could try and group average with respect to these
operators.

Another open issue pertains to the representation appropriate to the case of non- compact spatial topology.
The quantization here explicitly incorporates the compact spatial topology $S^1$. Here, the unit of length has been
chosen so that the circumference of the $T$ = constant circle is $2\pi$. By allowing the circle to have an 
arbitrarily large circumference, it may be possible to transit to polymer PFT on $R\times R$ and compare the 
resulting quantization with the Infinite Tensor Product proposal of Thiemann (\cite{ITP,ITP1}).

\section*{Appendix}
\section*{A.  Lemmas concerning Semiclassicality and Weaves.}

\noindent{\bf Lemma 1}: If $\Delta k_J \geq \pi$  (see (\ref{defki}),(\ref{defdelta})) for some
$J, \ 1\leq  J\leq N$ then $-1\leq f_{s,m=1}\leq \pi$.\\
\noindent{\bf Proof}:
Let $\Delta k_J \geq \pi$. Equations (\ref{sumdeltak}) imply that
\begin{equation}
\sum_{I\neq J}\Delta k_I\leq \pi,
\label{0.1}
\end{equation}
and, hence, that 
\begin{equation}
\Delta k_I|_{I\neq J} \leq \pi. 
\label{0.2}
\end{equation}
This
in conjunction with  the fact that 
$|\frac{\sin x}{x}|\leq 1$ implies that
\begin{equation}
\sum_{I=1}^N \sin \Delta k_I\leq \sum_{I\neq J} \Delta k_I + \sin \Delta k_J \leq \pi.
\label{upper}
\end{equation}
From equation (\ref{0.2}) and $\Delta k_J \geq \pi$, we have that 
\begin{equation}
\sum_{I=1}^N \sin \Delta k_I \geq -1 .
\label{lower}
\end{equation}
The Lemma follows immediately from equations (\ref{upper}), (\ref{lower})  and the definition (\ref{sindeltak})
of $f_{s,m=1}$

\noindent{\bf Lemma 2}: If $\Delta k_I \leq \pi, \ I=1,..,N$ (see (\ref{defki}),(\ref{defdelta}))
then $0\leq f_{s,m=1}\leq 2\pi$.\\
\noindent {\bf Proof}:  This follows immediately from the fact that $|\frac{\sin x}{x}|\leq 1$ in conjunction 
with equations (\ref{sumdeltak}) and the definition (\ref{sindeltak}) of $f_{s, m=1}$.

\noindent {\bf Lemma 3}: Equation (\ref{6.1.6}) 
implies that as $\epsilon \rightarrow 0$, 
$\Delta k_I \rightarrow 0 ,\; I= 1,..,N$ and $N\rightarrow \infty$.

\noindent{\bf Proof}: 

From Lemma 1 and equation (\ref{6.1.6}) it follows that for sufficiently small $\epsilon$, it must be the case
that $\Delta k_I \leq \pi, I=1,..,N$.

Next, let $\alpha$ be the minimum value of the bounded, continuous function $\frac{\sin \theta}{\theta}$ in the interval
$[0,\frac{\pi}{2}]$ (here $\frac{\sin \theta}{\theta}\vert_{\theta=0}:=1$). Define the function
$f(x):= x -\sin x -\frac{\alpha}{6}x^3$. It is easy to check that $\frac{df}{dx}\geq 0, \ x\in [0,\pi]$ and that
$f(x=0)=0$. This implies that $x- \sin x \geq \frac{\alpha}{6}x^3, \ x\in [0,\pi]$. This in conjunction with 
equations (\ref{sumdeltak}), (\ref{6.1.6}) implies that
$\sum_{I=1}^{N} (\Delta k_I)^3 < \frac{6 \epsilon}{\alpha}$ so that $\Delta k_I\rightarrow 0, \ I=1,..,N$ as 
$\epsilon \rightarrow 0$. This in turn, together with (\ref{sumdeltak}), implies 
that $N\rightarrow \infty$ as $\epsilon \rightarrow 0$.

\noindent{\bf Lemma 4}: Any normalised $|\psi \rangle \in {\cal H}_{kin}$ admits the expansion:
\begin{eqnarray}
|\psi \rangle &=& \sum_j a_j |s_j,\psi_{jM}\rangle, \;\;|s_j,\psi_{jM}\rangle:=
|s_j\rangle\otimes |\psi_{jM}\rangle , \label{lincomb}\\
\langle s_i|s_j\rangle &=&\delta{ij},\; s_j=\{\gamma (s_j), (k_{e^j_1},..,k_{e^j_{n_j}})\}\\
\langle \psi_{jM}|\psi_{jM}\rangle&=& 1, \label{normpsi}\\
\sum_{j} |a_j|^2 &=&1 .\label{norma}
\end{eqnarray}
Here $s_j$ are embedding charge labels, $e^j_I, I=1,.., n_j$ are the edges of the graph underlying $s_j$,
$a_j$ are complex coefficients and $|\psi_{jM}\rangle\in {\cal H}_M$.

If $|\psi \rangle$ is semiclassical then the coefficients  $a_j$ are such that 
$|\psi \rangle$ is peaked around $s_j$ such that $s_j$ are weaves.

\noindent{\bf Proof}: The proof closely mirrors the arguments of section 6.1. 
Semiclassicality implies that to leading order in $\hbar$,
\begin{equation}
\langle \psi |[\widehat{e^{i\alpha q_m}},\widehat{e^{i\alpha p_m}}]|\psi \rangle
\approx i\hbar \{e^{i \alpha q_m},e^{i \beta p_m}\}= -i\hbar \alpha \beta 2\pi m e^{i \alpha q_m +i \beta p_m}
\label{l2.1}
\end{equation}
Using  equations (\ref{lincomb}), (\ref{hateof}), (\ref{l2.1}), we have that
\begin{equation}
\sum_{j}|a_j|^2 2\sin (\frac{\alpha \beta \hbar}{2} f_{s_j, m})
\langle s_j, \psi_{jM} |\widehat{e^{i\alpha q_m+i\beta p_m}}|s_j, \psi_{jM} \rangle
\approx \hbar \alpha \beta 2\pi m e^{i \alpha q_m +i \beta p_m}
\label{l2.2}
\end{equation}
where 
\begin{equation}
f_{s_j,m} = \sum_{I=1}^{n_j} \sin m \Delta k^j_I,
\label{fsim}
\end{equation}
and $\Delta k^j_I := k^j_{I+1}-k^j_I$, $k^j_{n_j+1}:= k^j_{1}$ and we have set $k^j_I:= \hbar k_{e^j_I}$.

From Lemmas 1 and 2 it follows that 

\begin{equation}
-1\leq f_{s_j,m=1} \leq 2\pi .
\label{absfsj}
\end{equation}
Since $f_{s_j,m=1}$ is bounded, equation (\ref{l2.2}) implies that to leading order in $\hbar$, we have that 
\begin{equation}
\sum_{j}|a_j|^2 f_{s_j, m=1}\langle s_j, \psi_{jM} |\widehat{e^{i\alpha q_1+i\beta p_1}}|s_j, \psi_{jM} \rangle
e^{-i \alpha q_1 -i \beta p_1}\approx 2\pi
\label{l2.2.1}
\end{equation}
Denote the left hand-side of equation (\ref{l2.2.1}) by $LHS$. Equation (\ref{l2.2.1}) implies that 
\begin{equation}
|LHS -2\pi|\leq \delta, \;\;\; \delta <<1 .
\label{l2.2.3}
\end{equation}
Taking absolute values of both sides of equation (\ref{l2.2.1}) and
using (\ref{absfsj}), (\ref{norma}) and the fact that $\widehat{e^{i\alpha q_m+i\beta p_m}}$
is a bounded operator of norm 1, we have that 
\begin{equation}
2\pi \geq \sum_{j}|a_j|^2 |f_{s_j, m=1}|\geq |LHS|.
\label{l2.2.2}
\end{equation}
From (\ref{l2.2.2}), (\ref{l2.2.3}) we have that 
$\delta\geq |2\pi- LHS|\geq 2\pi -|LHS| \geq 2\pi - \sum_{j}|a_j|^2 |f_{s_j, m=1}|$,
so that 
\begin{equation}
\sum_{j}|a_j|^2 |f_{s_j, m=1}| \geq 2\pi -\delta .
\label{l2.2.4}
\end{equation}
Let $J_<$ be the set of all $j$ such that $|f_{s_j, m=1}| \leq 2\pi -\delta^{\frac{1}{2}}$ and
let $\sum_{j\in J_<}|a_j|^2 = P_<$. Then (\ref{absfsj}),(\ref{l2.2.4}) imply that 
$P_< (2\pi -\delta^{\frac{1}{2}})+ (1-P_<)2\pi \geq 2\pi - \delta$ so that $P_< \leq \delta^{\frac{1}{2}}$.
Thus as $\delta\rightarrow 0$, almost all $j$ are such that 
$|f_{s_j, m=1}| \geq 2\pi -\epsilon$ where we have set $\epsilon := \delta^{\frac{1}{2}}$. Using (\ref{absfsj}),
this, in turn, implies that for small enough $\epsilon$,
\begin{equation}
f_{s_j,m=1} \geq 2\pi - \epsilon.
\end{equation}
This brings us back to equation (\ref{6.1.5}) with $s=s_j, m=1$. The  analysis subsequent  to that equation implies that 
such $s_j$ must be a weave.

\noindent {\bf Lemma 5}: Let $|\psi\rangle\in {\cal H}_{phys}$ be semiclassical. Then $|\psi\rangle$ is peaked at 
group averages of embedding eigen states which are based on weaves.\\
\noindent{\bf Proof}: Recall that $|\psi\rangle$ is in the completion of $\eta ({\cal D})$ where ${\cal D}$ is the finite
span of charge network states. It is then straightforward to see that any such $|\psi\rangle$ admits the expansion:
\begin{equation}
|\psi \rangle = \sum_j a_j \eta (|s_j\rangle\otimes |\psi_{jM}\rangle) , \label{l3.1}
\end{equation}
such that 
\begin{equation}
\eta (|s_i\rangle\otimes |\psi_{iM}\rangle)[|s_j\rangle\otimes |\psi_{jM}\rangle] = \delta_{ij}, \label{l3.2}
\end{equation}
and $|s_i\rangle$, $|s_j\rangle$  are not gauge related if $i\neq j$ i.e. for $i\neq j$ and $\forall \phi$,  
\begin{equation}
|s_i\rangle \neq {\hat U}(\phi )|s_j\rangle .
\label{l3.3}
\end{equation}
Here $s_j$ is an embedding charge network label, $\phi$ is a gauge transformation and $|\psi_{jM}\rangle \in {\cal H}_M$.
We shall use the notation of Lemma 4 for the edges and charge labels of $s_j$. Note that $|\psi_{jM}\rangle$ is
such that $\eta (|s_j\rangle\otimes |\psi_{jM}\rangle ) \in {\cal H}_{phys}$ as implied by (\ref{l3.2}).
Using (\ref{physip}), the normalization $<\psi|\psi>_{phys}=1$ implies that 
\begin{equation}
\sum_{j}|a_j|^2 =1
\label{l3.4}
\end{equation}
Semiclassicality implies that, to leading order in $\hbar$,
\begin{equation}
\langle \psi |[\widehat{e^{i\alpha q_m}},\widehat{e^{i\alpha p_m}}]|\psi \rangle_{phys}
\approx -i\hbar \alpha \beta 2\pi m e^{i \alpha q_m +i \beta p_m},
\label{l3.5}
\end{equation}
where the `-' sign  in the right hand side is due to the fact that operators act on ${\cal H}_{phys}$  by dual action
(see Footnote \ref{dualaction}). Using equations (\ref{hateof}) and  (\ref{l3.3}), we have that
\begin{eqnarray}
&&-\sum_{j}|a_j|^2 2i\sin (\frac{\alpha \beta \hbar}{2} f_{s_j, m})
\langle \eta (|s_j\rangle \otimes |\psi_{jM}\rangle),\  \widehat{e^{i\alpha q_m+i\beta p_m}}
\eta (|s_j\rangle\otimes | \psi_{jM} \rangle)\rangle_{phys}\nonumber \\  
&\approx & 
-i\hbar \alpha \beta 2\pi m e^{i \alpha q_m +i \beta p_m}.
\label{l3.6}
\end{eqnarray}
Here $f_{s_j,m}$ is defined as in Lemma 4.
\footnote{It is straightforward to check that $f(s_j,m)$ (\ref{fsim}) is a {\em gauge invariant} function
of $s_j$ i.e. $f_{s_j,m}= f (s^{\prime}_j, m) \forall s^{\prime}_j$ such that $\exists$ a gauge transformation
$\phi$ such that $|s^{\prime}_j\rangle = {\hat U}(\phi )|s_j\rangle$.}
This is the analog of equation (\ref{l2.2}) of Lemma 4. 
The analysis of Lemma 4 subsequent to that 
equation applies here identically thus proving Lemma 5.

\section*{B. Lemmas concerning the no go result of section 6.2.}

\noindent {\bf Lemma 6}: 
No states $|\psi\rangle \in {\cal H}_{kin}$ exist which are semiclassical with respect to the uncountable set of 
operators $\{\widehat{e^{i\alpha q_m}},\widehat{e^{i\beta p_m}}, |\alpha -\alpha_0|<\epsilon,
|\beta - \beta_0|<\delta\}$ for any fixed $m, \alpha_0, \beta_0$ and any $\epsilon, \delta >0$.

\noindent{\bf Proof}: As in Lemma 4 of Appendix A, any
$|\psi\rangle \in {\cal H}_{kin}$ admits the expansion (\ref{lincomb})- (\ref{norma}). Additionally
we may expand $|\psi_{jM}\rangle$ in terms of matter charge networks so that for any fixed $j$,
\begin{eqnarray}
|\psi_{jM}\rangle & = &\sum_{r^j}b_{r^j}|s^{\prime }_{r^j}\rangle \label{l6.1}\\
\langle s^{\prime }_{r^j_1}|s^{\prime }_{r^j_2}\rangle = \delta_{r^j_1, r^j_2}\label{6.2}
\end{eqnarray}
where $r^j$ varies over a countable set (as, of course, does $j$), $b_{r^j}$ are complex coefficients and 
and $s^{\prime}_{r^j}$ are matter charge networks.

Let $C$ be the set of all $j$ such that $\gamma (s_j)$ has at least one edge $e(j)$ with embedding 
charge $k_{e(j)}$ such that 
$\cos m\hbar k_{e(j)} \neq 0$. For every $j\in C$ choose  an edge $e^j\subset \gamma (s_j)$ with embedding 
charge $k_{e^j}$
such that 
\begin{equation}
c_j := \cos m\hbar k_{e(j)} \neq 0. 
\label{cj}
\end{equation}
Let $S$ be the set of all $j$ such that $j\notin C$. Clearly, for each
$j\in S$ we can fix an edge $e^j\in \gamma (s_j)$ such that its charge label $k_{e^j}$ satisfies 
\begin{equation}
s_j:= \sin m\hbar k_{e(j)} \neq 0.
\label{sj}
\end{equation}
Next, let $L$ be the set of all matter charges which occur in $s^{\prime}_{r^j} \forall j,r$. Let $\Delta L$ be
the set of differences between all pairs of elements of $L$ i.e. 
$\Delta L = \{ l-l^{\prime} \forall l,l^{\prime} \in L\}$. For every $j_C\in C$, $j_S\in S$, define the 
sets 
$\Delta L_{j_C}, \Delta L_{j_S}$ whose elements are obtained by dividing elements of $\Delta L$ by $c_{j_C},s_{j_S}$
(see (\ref{cj}),(\ref{sj})) i.e.
$\Delta L_{j_C}:= \{\frac{x}{c_{j_C}} \forall x \in \Delta L\}$,
$\Delta L_{j_S}:= \{\frac{x}{s_{j_S}} \forall x \in \Delta L\}$.
Finally, let $\Delta L_C:= \cup_{j_C\in C} L_{j_C},\Delta L_S:= \cup_{j_S\in S} L_{j_S}$.

Note that $\Delta L_C, \Delta L_S$ are both countable sets.
It follows that in any neighbourhood of $\alpha_0, \beta_0$ there exist uncountably many $\alpha, \beta$ such that
$\alpha \notin  \Delta L_C, \beta \notin \Delta L_S$.  Then from (\ref{hateof}) and the fact that 
$\widehat{e^{i\beta p_m}}$ is an operator of unit norm, it follows that for such $\alpha, \beta$, 
\begin{equation}
|\langle \psi \widehat{e^{i\alpha q_m}}|\psi\rangle |= \sum_{j\in S}|a_j|^2,
\label{6big1}
\end{equation}
\begin{equation}
|\langle \psi \widehat{e^{i\beta p_m}}|\psi\rangle| \leq \sum_{j\in C}|a_j|^2 = 1 - \sum_{j\in S}|a_j|^2 .
\label{6big2}
\end{equation}
Semiclassicality requires that both
(\ref{6big1}) and (\ref{6big2}) be close to unity. Clearly, this is not possible.\\

\noindent{\bf Lemma 7}: 
No states $|\psi\rangle \in {\cal H}_{phys}$ exist which are semiclassical with respect to the uncountable set of 
operators $\{\widehat{e^{i\alpha q_m}},\widehat{e^{i\beta p_m}}, |\alpha -\alpha_0|<\epsilon,
|\beta - \beta_0|<\delta\}$ for any fixed $m, \alpha_0, \beta_0$ and any $\epsilon, \delta >0$.\\

\noindent{\bf Proof}: As in Lemma 5, Appendix A, any $|\psi\rangle \in {\cal H}_{phys}$ admits the expansion
(\ref{l3.1})- (\ref{l3.3}). Further 
$|\psi_{jM}\rangle$ can be expanded as in equation (\ref{l6.1})- (\ref{6.2}) of Lemma 6.
Note that the antilinearity of $\eta$ implies that we may rewrite equation
(\ref{l3.1}) as
\begin{equation}
|\psi \rangle = \eta (\sum_j a^*_j |s_j\rangle\otimes |\psi_{jM}\rangle) . \label{l7.1}
\end{equation}
Next, let us construct the sets $\Delta L_C, \Delta L_S$ (as defined in Lemma 6) for the state 
$\sum_j a^*_j |s_j\rangle\otimes |\psi_{jM}\rangle \in {\cal H}_{kin}$. 
It follows straightforwardly from the periodicity of the cosine and sine functions in conjunction with the 
action of gauge transformations (\ref{bfsphi}) that we may choose the sets $\Delta L_C, \Delta L_S$ in such a way
that they are identical for
any (kinematic) state which is gauge related to the state
$\sum_j a^*_j |s_j\rangle\otimes |\psi_{jM}\rangle$. Thus the sets $\Delta L_C, \Delta L_S$ can be chosen so as to
depend only on the physical state $|\psi \rangle$, and it is straightforward to see that, as in Lemma 6,
if we choose $\alpha \notin \Delta L_C$,$\beta \notin \Delta L_S$, we obtain equations (\ref{6big1}), (\ref{6big2})
with $|\psi \rangle$ as in (\ref{l7.1}). This proves the Lemma.
\\
\section*{C.  Choice of units.}
\hspace*{0.2in} In this appendix we summarize dimensions of various operators and parameters of the theory. We have set
the speed of light $c$ to be unity.
\begin{equation}
\begin{array}{lll}
[S_{0}] = ML = [\hbar]\\
\vspace*{0.1in}
[f] = M^{\frac{1}{2}}L^{\frac{1}{2}},\ [\pi_{f}] = M^{\frac{1}{2}}L^{-\frac{1}{2}}\\
\vspace*{0.1in}
[X^{\pm}] = L, [\Pi_{\pm}] = ML^{-1}\\
\vspace*{0.1in}
[q_{(\pm)n}] = M^{\frac{1}{2}}L^{-\frac{1}{2}} = [p_{(\pm)n}]
\end{array}
\end{equation}
where $[n] = L^{-1}$.\\
\hspace*{0.2in} The dimensions of the above fields naturally imply the dimensions of the various charges and parameters involved in the theory.
\begin{equation}
\begin{array}{lll}
[k_{e}] = M^{-1},\ [l_{e}] = M^{-\frac{1}{2}}L^{-\frac{1}{2}}\\
\vspace*{0.1in}
[\alpha] = M^{-\frac{1}{2}}L^{-\frac{1}{2}}
\end{array}
\end{equation}
where the parameter $\alpha$ occurs in the exponentiated observables defined in (\ref{eq:81}).\\
\hspace*{0.2in} Throughout this paper, we have fixed the units such that length of the $T$= constant circle 
is $2\pi$. Thus the only arbitrary scale in the theory is the mass scale.

\end{document}